\documentclass[pdflatex,sn-mathphys-num,iicol]{sn-jnl}

\usepackage{amsmath,amssymb,mathtools}
\usepackage{bm}
\usepackage{graphicx}
\usepackage{booktabs}
\usepackage{array}
\usepackage{multirow}
\usepackage{makecell}
\usepackage{siunitx}
\usepackage{xcolor}
\usepackage{adjustbox}
\usepackage{microtype}

\hypersetup{
  pdfnewwindow=true,
  colorlinks=true,
  linkcolor=blue,
  citecolor=blue,
  filecolor=blue,
  urlcolor=blue
}

\allowdisplaybreaks[2]
\newcommand{\ba}{\begin{eqnarray}}
\newcommand{\ea}{\end{eqnarray}}

\newif\ifuseprebuiltbbl
\useprebuiltbbltrue

\begin{document}

\title[Strong decay widths of heavy baryons]
{Strong decay widths of $S$- and $P$-wave singly- and doubly-heavy charm and bottom baryons}

\author*[1,2]{\fnm{Emmanuel} \sur{Ortiz-Pacheco}}
\email{emmanuel.ortizpac@gmail.com}

\author[3]{\fnm{Roelof} \sur{Bijker}}
\email{bijker@nucleares.unam.mx}

\affil[1]{
  \orgdiv{Institute for Theoretical Physics},
  \orgname{ELTE E\"otv\"os Lor\'and University},
  \orgaddress{
    \street{P\'azm\'any P\'eter s\'et\'any 1/A},
    \city{Budapest},
    \postcode{H-1117},
    \country{Hungary}
  }
}

\affil[2]{
  \orgname{MTA--ELTE Lend\"ulet ``Momentum'' Strongly Interacting Matter Research Group},
  \orgaddress{
    \city{Budapest},
    \country{Hungary}
  }
}

\affil[3]{
  \orgdiv{Instituto de Ciencias Nucleares},
  \orgname{Universidad Nacional Aut\'onoma de M\'exico},
  \orgaddress{
    \city{Ciudad de M\'exico},
    \postcode{04510},
    \country{Mexico}
  }
}


\abstract{We present a study of two-body decay widths of heavy baryons into another heavy baryon and a light pseudoscalar meson in the framework of the non-relativistic quark model in combination with the elementary emission model for the strong couplings. The present study includes the decays of $1S$- and $1P$-wave baryons with one or two heavy quarks, either charm ($c$) or bottom ($b$). The flavor, spin and orbital contributions are discussed explicitly as well as selection rules for forbidden decays. The relative partial widths are given by the appropriate flavor $SU(3)$ isoscalar factors. The total widths are compared with the available experimental data as well as with other theoretical studies. The calculated widths of singly-heavy baryons are found to be in reasonable agreement with the observed widths. The decay widths of doubly-heavy baryons are suppressed with respect to those of the singly-heavy baryons by the ratios of the quark masses appearing in the orbital contributions.}


\maketitle

\section{Introduction}
\label{sec:introduction}

The discovery of many new hadrons with heavy quarks by the LHCb, CMS, CDF, Belle and BaBar Collaborations has stimulated an enormous activity in the field of hadronic physics both experimentally and theoretically \cite{Chen_2023}. At the LHC alone 85 new hadrons were reported since 2012, divided between 36 baryons, 25 mesons and 24 exotics \cite{koppenburg_particles}. The study of heavy baryons and mesons provides an important tool to improve our understanding of hadron structure, the dynamics of the strong interaction and heavy-quark symmetry. 

Baryons with a single heavy quark, either charm $c$ or bottom $b$, belong to a $SU(3)$ flavor sextet or a flavor anti-triplet. The ground state baryons of the sextet $\Sigma_Q$, $\Xi'_Q$ and $\Omega_Q$ have spin and parity $J^P=1/2^+$ or $3/2^+$, and those of the anti-triplet $\Lambda_Q$ and $\Xi_Q$ have $J^P=1/2^+$. In the charm sector all of these states have been measured, and in the bottom sector all with the exception of only one, $\Omega_b$ with $J^P=3/2^+$ \cite{PDG26}. 

In addition, in recent years the spectroscopy of single heavy baryons was enriched by the discovery of many orbitally excited states, both in the charm sector: $\Sigma_c$  \cite{PhysRevD.84.012003,PhysRevLett.94.122002,PhysRevD.104.052003,lhcbcollaboration2026observationnewexciteditsigmac0}, $\Lambda_c$    
\cite{PhysRevD.84.012003,JHEP05(2017)030,PhysRevLett.98.012001,PhysRevLett.98.262001}, $\Xi'_c$ and $\Xi_c$  \cite{PhysRevLett.124.222001,PhysRevD.102.071103,MoonPRD103,PhysRevD.94.032002,PhysRevD.94.052011,PhysRevD.89.052003,gghl-m6fm,PhysRevLett.134.081901,62lj-n7qw,luo2026quantumnumbersexcitedxicprime} and $\Omega_c$  \cite{PhysRevLett.131.131902,PhysRevLett.118.182001,PhysRevD.97.051102,PhysRevD.104.L091102}, and in the bottom sector: $\Sigma_b$ \cite{PhysRevLett.122.012001,PhysRevD.85.092011,PhysRevLett.99.202001}, $\Lambda_b$  \cite{PhysRevLett.109.172003,lamb,PhysRevLett.123.152001,PhysRevD.88.071101,PLB803135345}, $\Xi'_b$ and $\Xi_b$ \cite{PhysRevLett.126.252003,PhysRevLett.121.072002,PhysRevD.103.012004,lhcbcollaboration2023observation} and $\Omega_b$  \cite{PhysRevLett.124.082002}. Most excited baryons were interpreted as $1P$-wave states with one quantum of excitation, and some as $2S$- or $1D$-wave states with two quanta of excitation. 

Baryons with two heavy quarks form two flavor triplets, $\Xi_{QQ}$ and $\Omega_{QQ}$, with spin and parity $J^P=1/2^+$ and $3/2^+$. Candidates for all three members of the triplet with $J^P=1/2^+$ have been identified experimentally~\cite{Wang:2026OmegaCC1,LHCb:2026OmegaCC2}. The LHCb Collaboration found a signal for the $\Xi^{++}_{cc}(3621)$ ($ucc$) baryon in the $\Lambda_c^+ K^-\pi^+\pi^+$ invariant mass \cite{PhysRevLett.119.112001}. 
More recently, the isospin partner $\Xi^{+}_{cc}$ ($dcc$) has been discovered by the LHCb Collaboration with a mass $3620^{+2.1}_{-1.6}$ (MeV) \cite{dmv6-7gdv}.
Earlier evidence for the existence of a double charm baryon reported by the SELEX Collaboration, $\Xi^{+}_{cc}(3519)$ \cite{PhysRevLett.89.112001,OCHERASHVILI200518}, could not be confirmed by the BaBar, Belle and LHCb collaborations \cite{Xicc2019}. As a consequence, $\Xi^{+}_{cc}(3519)$ has been omitted from the Particle Data Group Summary Table \cite{PDG26}. 
At the Beauty 2026 conference celebrated in Maastricht there was the first announcement of the discovery of the doubly-charmed baryon $\Omega_{cc}^{+}$ ($scc$) in the
$\Omega_{c}^{0}\pi^{+}$ invariant-mass spectrum  with a mass $3727 \pm 1.2$ (MeV)~\cite{Wang:2026OmegaCC1,LHCb:2026OmegaCC2}.

The physics of heavy baryons has been studied theoretically in a vast variety of approaches such as
non-relativistic quark models (NRQM) \cite{Valcarce2008,PhysRevD.90.094007,PhysRevD.92.114029,PhysRevD.101.074031,doi:10.1142/S0217751X22502256,Chen2017,Santopinto2019,PhysRevD.105.074029,PhysRevD.107.034031,garciatecocoatzi2023decay}, relativistic quark models (RQM) \cite{PhysRevD.66.014008,PhysRevD.84.014025,YU2023116183,10.1142/S0217751X23500823,Migura2006},  hypercentral quark models (hCQM) \cite{Shah_2016,Shah2016epjc,Shah2017epja,Gandhi2018plus,Gandhi2020,Kakadiya2023,Mutuk_2020}, chiral quark models ($\chi$QM) \cite{PhysRevD.96.094005,WangOme,Wang2017,PhysRevD.103.074025}, Regge phenomenology \cite{PhysRevD.95.116005,Song2023,oudichhya2023}, QCD sum rules (QCDSR) \cite{Aliev2009,Aliev2015,Aliev2019,Aliev2016,PhysRevD.102.114009,PhysRevD.91.054034,PhysRevD.101.114013,PhysRevD.104.034037}, Faddeev methods \cite{Garcilazo_2007,PhysRevD.100.034008}, molecular states \cite{PhysRevD.101.054033,Zhu2020}, large $N_c$ limit \cite{YANG2020135142}, heavy baryon chiral perturbation theory (HB$\chi$PT) \cite{JuanWang2019}, quark-diquark model \cite{Mutuk_2021,PhysRevD.107.074015} and lattice QCD calculations (LQCD) \cite{MeinelPRD85,PhysRevD.90.074504,PhysRevD.87.094512,PhysRevD.90.074501,PhysRevD.91.094502,PhysRevD.92.034504,PhysRevD.96.034511,PhysRevD.90.094507,PhysRevD.86.094504}. Reviews of heavy baryon physics together with a more complete list of references can be found in Refs.~\cite{KORNER1994787,Roberts2008,VIJANDE2013,rev2015,RPP76,Workman:2022ynf,Chen_2023,Crede:2026had}.

The aim of this article is to present a simultaneous study of strong couplings of all $1S$- and $1P$-wave baryons with one or two heavy quarks, either charm ($c$) or bottom ($b$). This work is a follow-up of previous studies of $\Omega_Q$ baryons \cite{Santopinto2019,Ortiz-Pacheco:2020hmj}, $\Xi'_Q$ and $\Xi_Q$ baryons \cite{PhysRevD.105.074029}, and a simultaneous analysis of mass spectra and electromagnetic couplings of all $1S$- and $1P$-wave heavy baryons \cite{Ortiz-Pacheco:2021thesis,EOP2023} including $\Sigma_Q$, $\Lambda_Q$, $\Xi_{QQ}$ and $\Omega_{QQ}$ baryons as well \cite{ncc47175}. 

This contribution is organized as follows. In Section~\ref{strong} we discuss the strong couplings in the elementary emission model (EEM). The spin-flavor matrix elements are evaluated using the $SU(3)$ flavor symmetry with the phase convention of Baird and Biedenharn \cite{Baird:1963wv}. The necessary radial integrals and flavor wave functions are presented in the Appendices \ref{app:radial}, \ref{app:wf} and \ref{app:meson}. In section~\ref{widths}, we discuss selection rules for strong decays and present the two-body strong decay widths of single and double heavy baryons. Finally, in Section~\ref{discussion} we present a comparison with other model predictions and in Section~\ref{summary} we finish with a brief summary and conclusions. 

\section{Strong couplings}
\label{strong}

\begin{figure}[t]
\centering
\includegraphics[width=\columnwidth]{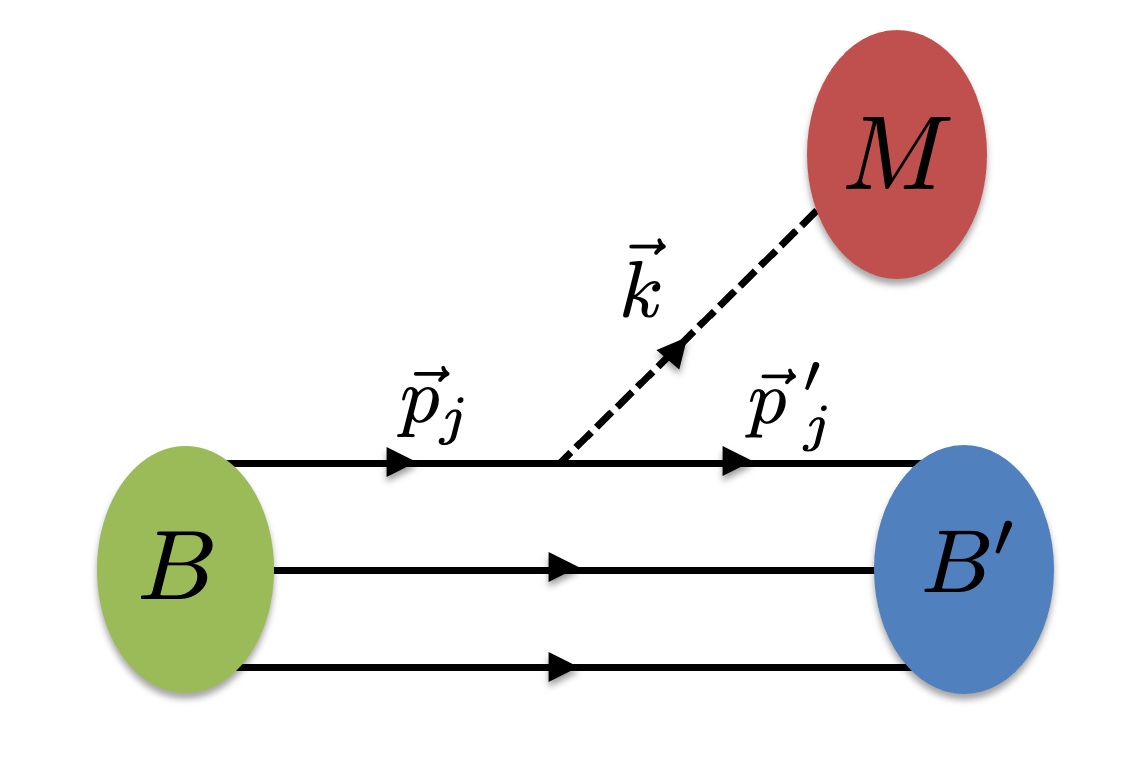} 
\caption[]{Strong decay diagram of a baryon $B$ decaying into another baryon $B'$ with the emission of a meson  $M$
: $B\rightarrow B'+M$.} 
\label{BBM}
\end{figure}

Strong couplings provide an important test of the structure of baryons. Given the intrinsically non-perturbative nature of QCD, the study of strong decays presents a significant challenge. Heavy baryons exhibit a broad range of strong decay channels, including fully heavy and open-flavor  baryons. Among these, an important class consists of two-body decays, where a baryon decays into two hadrons. In particular, baryon to baryon-meson transitions are of great interest. In this work, the elementary emission model (EEM) is employed to study such decays \cite{LeYaouanc:1988fx}. Within this framework, a baryon, treated as a composite system of quarks, emits a meson through one of its constituent quarks, see Fig.~\ref{BBM}. The meson is treated as a fully formed hadron, and the total number of quarks remains conserved before and after the emission. The strong interaction of the meson emission, like the fundamental interactions of QCD between fermions and gauge bosons, is a three-line interaction between quarks and the elementary meson as in Fig.~\ref{BBM}. In the present article, we use the EEM to study the decay of heavy baryons $B_Q \rightarrow B'_Q + M$ into another heavy baryon and a light pseudoscalar meson, $\pi$, $K$, $\eta$, and $\eta'$

The strong interaction Hamiltonian to study the baryon  decays with meson emission can be obtained by introducing a vector-axial coupling through a QCD Lagrangian density that considers the interaction between the axial current of quarks and the derivative of the pseudoscalar meson field \cite{LeYaouanc:1988fx}
\ba
\mathcal{L}_I(x)=-\frac{g_{qq'M}}{2f_M}\bar{q}(x)\gamma^\mu\gamma_5\tau^a q(x)\partial_\mu\varphi^a(x),
\ea
with $g_{qq'M}$ the coupling constant and $f_M$ the meson decay constant.
The resolution into constituent quarks naturally restricts our study to a non-relativistic approximation, where the interaction Hamiltonian is given by
\cite{LeYaouanc:1988fx,CAPSTICK2000S241,PhysRevD.21.1868,BIJKER200089,PhysRevD.103.094003}
\begin{equation}
\begin{aligned}
\mathcal H_s
 &= \frac{1}{(2\pi)^{3/2}\sqrt{2k_0}}
 \sum_{j=1}^{3} X_j^M
 \Bigl[
 2g(\vec s_j\!\cdot\!\vec k)
 e^{-i\vec k\cdot\vec r_j}
 \\[-1mm]
 &\qquad
 +h\,\vec s_j\!\cdot\!
 \bigl(
 \vec p_j e^{-i\vec k\cdot\vec r_j}
 +e^{-i\vec k\cdot\vec r_j}\vec p_j
 \bigr)
 \Bigr].
\end{aligned}
\label{hs}
\end{equation}
Here $X_j^M$ is the flavor operator for the emission of a meson $M$ from the $j$-th quark, $q_j \rightarrow q'_j + M$, and $\vec{s_j}$, $\vec{r_j}$ and $\vec{p_j}$ denote the spin, coordinate and momentum of the $j$-th constituent quark, respectively. The meson energy is denoted by $k_0=E_M=E_{B}-E_{B'}$ and the meson momentum by $\vec{k}=\vec{P}-\vec{P}'=\vec{p}_j-\vec{p}_j^{\, '}$. The strong coupling constants, $g$ and $h$, are determined in a global fit to the available baryon decay widths from experimental data as discussed in more detail in Section~\ref{widths}.

The decay widths are calculated in the rest frame of the initial baryon $B$ and the $z$-axis is chosen in the direction of the meson momentum, $\vec{k}=k\hat{z}$. First, we make a transformation to center-of-mass and Jacobi coordinates 
\ba
\vec \rho    &=& \frac{1}{\sqrt{2}} (\vec r_1 - \vec r_2) ,
\nonumber\\
\vec \lambda &=& \frac{1}{\sqrt{6}} (\vec r_1 + \vec r_2 - 2 \vec r_3) ,
\nonumber\\
\vec R   &=& \frac{m(\vec r_1 + \vec r_2) + m'\vec r_3}{2m+m'} ,
\label{jacobi}
\ea 
with $m_1=m_2=m \neq m'=m_3$, and their conjugate momenta, followed by an integration over the center-of-mass coordinate to obtain a simplified form of the operator \cite{BIJKER200089} 
\begin{equation}
\begin{aligned}
\mathcal H_s
 &= \frac{1}{(2\pi)^{3/2}\sqrt{2k_0}}
 \notag\\[-1mm]
 &\quad\times
 \sum_{j=1}^{3} X_j^M
 \Biggl[
 \left(2gk-\frac{m_j}{M}hk\right)
 s_{j,z}\hat U_j
 \\[-1mm]
 &\qquad
 +2h\,s_{j,z}\hat T_{j,z}
 \\[-1mm]
 &\qquad
 +h\left(
 s_{j,+}\hat T_{j,-}
 +s_{j,-}\hat T_{j,+}
 \right)
 \Biggr].
\end{aligned}
\label{hstrong}
\end{equation}
The explicit form of the operators, $\hat U$ and $\hat T$, is given by 
\ba
\hat U_1 &=& \exp\left[-ik \left( \frac{1}{\sqrt 2} \rho_z + \frac{1}{\sqrt 6} \frac{3m'}{2m+m'} \lambda_z \right) \right] 
\nonumber\\
\hat U_2 &=& \exp\left[-ik \left(-\frac{1}{\sqrt 2} \rho_z + \frac{1}{\sqrt 6} \frac{3m'}{2m+m'} \lambda_z \right) \right] 
\nonumber\\
\hat U_3 &=& \exp\left[-ik \left(-\sqrt{\frac{2}{3}} \frac{3m}{2m+m'} \lambda_z \right) \right] 
\ea
and
\begin{align}
\hat T_{1,\mu}
 &= \frac12\Biggl[
 \left(
 \frac{p_{\rho,\mu}}{\sqrt2}
 +\frac{p_{\lambda,\mu}}{\sqrt6}
 \right)\hat U_1
 \notag\\[-1mm]
 &\qquad
 +\hat U_1\left(
 \frac{p_{\rho,\mu}}{\sqrt2}
 +\frac{p_{\lambda,\mu}}{\sqrt6}
 \right)
 \Biggr],
 \notag\\
\hat T_{2,\mu}
 &= \frac12\Biggl[
 \left(
 -\frac{p_{\rho,\mu}}{\sqrt2}
 +\frac{p_{\lambda,\mu}}{\sqrt6}
 \right)\hat U_2
 \notag\\[-1mm]
 &\qquad
 +\hat U_2\left(
 -\frac{p_{\rho,\mu}}{\sqrt2}
 +\frac{p_{\lambda,\mu}}{\sqrt6}
 \right)
 \Biggr],
 \notag\\
\hat T_{3,\mu}
 &= -\frac12\sqrt{\frac23}
 \left(
 p_{\lambda,\mu}\hat U_3
 +\hat U_3p_{\lambda,\mu}
 \right).
\end{align}
The helicity amplitude for a two-body strong decay $B \rightarrow B' + M$ into a ground-state baryon $B'$ with $L'=0$, $J'=S'$ and a pseudoscalar meson $M$ is given by
\begin{align}
\mathcal A_\nu(k)
 &= \langle \psi_{B'};J',\nu
 \mid \mathcal H_s
 \mid \psi_B;J,\nu\rangle
 \notag\\
 &= \frac{1}{(2\pi)^{3/2}\sqrt{2k_0}}
 \Biggl\{
 \langle L,0,S,\nu\mid J,\nu\rangle
 \notag\\[-1mm]
 &\qquad\times
 \sum_{j=1}^{3}\zeta_{j,0}Z_{j,0}(k)
 \notag\\
 &\quad+\frac12
 \langle L,1,S,\nu-1\mid J,\nu\rangle
 \notag\\[-1mm]
 &\qquad\times
 \sum_{j=1}^{3}\zeta_{j,+}Z_{j,-}(k)
 \notag\\
 &\quad+\frac12
 \langle L,-1,S,\nu+1\mid J,\nu\rangle
 \notag\\[-1mm]
 &\qquad\times
 \sum_{j=1}^{3}\zeta_{j,-}Z_{j,+}(k)
 \Biggr\}.
\label{anu}
\end{align}
for helicities $\nu=1/2$, $3/2$ and 
\ba
Z_{j,0}(k) &=& \left(2gk-\frac{m_j}{M}hk\right)U^\alpha_j(k)+2h T^\alpha_{j,0}(k) ~,
\nonumber\\
Z_{j,\pm}(k) &=& 2h T^\alpha_{j,\pm}(k) ~.
\label{zeta}
\ea
The coefficients $Z_{j,\mu}(k)$ depend on the radial integrals, $U^\alpha_{j}$ and $T^\alpha_{j,\mu}$, which can be obtained in closed analytic form for the harmonic oscillator quark model (see Appendix~\ref{app:radial}). 

For three identical quarks, the contribution of each one of the quarks is the same and one can write
\ba
\mathcal{A}_\nu(k) &=& \langle \psi_{B'};J',\nu \mid \sum_{j=1}^3 {\cal H}_{s,j} 
\mid \psi_B;J,\nu \rangle 
\nonumber\\
&=& \langle \psi_{B'},J',\nu \mid 3{\cal H}_{s,3} \mid \psi_{B},J,\nu \rangle ~,
\ea
whereas for two identical quarks which are different from the third, one has
\ba
\mathcal{A}_\nu = \langle \psi_{B'},J',\nu \mid 2{\cal H}_{s,1} + {\cal H}_{s,3} 
\mid \psi_{B},J,\nu \rangle ~. 
\label{qqQ}
\ea

\subsection{Spin-flavor matrix elements}

\begin{figure}[t]
\centering
\setlength{\unitlength}{0.7pt}
\begin{picture}(130,160)(0,40)
\thicklines
\put( 50, 50) {\line(0,1){100}}
\put( 50, 50) {\vector(0,1){35}}
\put( 50,100) {\vector(0,1){35}}
\put( 50,100) {\circle*{5}}
\multiput(50,100)(5,5){10}{\circle*{2}}
\put( 35, 55) {$q$}
\put( 35,135) {$q'$}
\put( 80,110) {$q\bar{q}'$}
\end{picture}
\caption{Schematic representation of a meson emitted by light quark withing the EEM.}
\label{qqM}
\end{figure}

The coefficients $\zeta_j$ in Eq.~(\ref{anu}) correspond to the spin-flavor matrix elements 
\ba
\zeta_{j,\mu} &=& \langle \psi_{B'} | X_j^M s_{j,\mu} | \psi_{B} \rangle ~.
\ea
The flavor matrix elements can be evaluated by using the Wigner-Eckart theorem and isoscalar factors of the $SU(3)$ flavor group. The flavor wave functions are labeled by $(p,q)$ which are related to the Young tableau $[f_1f_2f_3]$ by $(p,q)=(f_1-f_2,f_2-f_3)$, isospin $I$ and hypercharge $Y$. For a given isospin channel the flavor matrix element can be expressed as \cite{BIJKER200089}
\begin{align}
\langle \phi_{B'}|X_j^M|\phi_B\rangle
 &= \left\langle
 \begin{matrix}(p',q')\\ I',Y'\end{matrix}
 \middle|
 T_j^{(p'',q''),I'',Y''}
 \middle|
 \begin{matrix}(p,q)\\ I,Y\end{matrix}
 \right\rangle
 \notag\\
 &= \sum_\gamma
 \left\langle
 \begin{array}{cc|c}
 (p',q')&(p'',q'')&(p,q)\\
 I',Y'&I'',Y''&I,Y
 \end{array}
 \right\rangle_\gamma
 \notag\\[-1mm]
 &\qquad\times
 \langle(p',q')\parallel
 T_j^{(p'',q'')}\parallel(p,q)\rangle_\gamma.
\label{fme}
\end{align}
The first term in the r.h.s. of Eq.~(\ref{fme}) is a $SU(3)$ isoscalar factor and the second term a $SU(3)$ reduced matrix element. The initial baryon $B$ is characterized by $(p,q)$, the final baryon $B'$ by $(p',q')$ and the flavor operator $X^M$ by $(p'',q'')$. The sum over $\gamma$ is over different multiplicities. In the present case it can be omitted since for strong decays of heavy baryons considered in this paper there are no multiplicities. Specific charge channels are obtained by multiplying with the appropriate isospin Clebsch-Gordan coefficient $\langle I',I'_3,I'',I''_3 | I,I_3 \rangle$.

\begin{figure}[t]
\centering
\setlength{\unitlength}{0.38pt}
\begin{picture}(550,1000)(-50,-125)
\thicklines
\multiput(150,750)(100,0){2}{\circle*{20}}
\multiput(100,700)(100,0){3}{\circle*{20}}
\multiput(150,650)(100,0){2}{\circle*{20}}
\put(150,750) {\line(1,0){100}}
\put(100,700) {\line(1,0){200}}
\put(150,650) {\line(1,0){100}}
\put(100,700) {\line(1,1){ 50}}
\put(150,650) {\line(1,1){100}}
\put(250,650) {\line(1,1){ 50}}
\put(300,700) {\line(-1,1){ 50}}
\put(250,650) {\line(-1,1){100}}
\put(150,650) {\line(-1,1){ 50}}
\put(375, 740) {$N(nnn)$}
\put(375, 690) {$\Sigma, \Lambda(nns)$}
\put(375, 640) {$\Xi(nss)$}
\put(-50, 690) {$(1,1) \equiv {\bf 8}$}

\multiput( 50,550)(100,0){4}{\circle*{20}}
\multiput(100,500)(100,0){3}{\circle*{20}}
\multiput(150,450)(100,0){2}{\circle*{20}}
\put(200,400){\circle*{20}}
\put( 50,550) {\line(1,0){300}}
\put(100,500) {\line(1,0){200}}
\put(150,450) {\line(1,0){100}}
\put(100,500) {\line(1,1){ 50}}
\put(150,450) {\line(1,1){100}}
\put(200,400) {\line(1,1){150}}
\put(300,500) {\line(-1,1){ 50}}
\put(250,450) {\line(-1,1){100}}
\put(200,400) {\line(-1,1){150}}
\put(375, 540) {$\Delta(nnn)$}
\put(375, 490) {$\Sigma^\ast(nns)$}
\put(375, 440) {$\Xi^\ast(nss)$}
\put(375, 390) {$\Omega(sss)$}
\put(-50, 440) {$(3,0) \equiv {\bf 10}$}
\multiput(100,300)(100,0){3}{\circle*{20}}
\multiput(150,250)(100,0){2}{\circle*{20}}
\put(200,200){\circle*{20}}
\put(100,300) {\line(1,0){200}}
\put(150,250) {\line(1,0){100}}
\put(200,200) {\line(1,1){100}}
\put(150,250) {\line(1,1){ 50}}
\put(200,200) {\line(-1,1){100}}
\put(250,250) {\line(-1,1){ 50}}
\put(375, 290) {$\Sigma_Q(nnQ)$}
\put(375, 240) {$\Xi'_Q(nsQ)$}
\put(375, 190) {$\Omega_Q(ssQ)$}
\put(-50, 240) {$(2,0) \equiv {\bf 6}$}
\put(200,100) {\circle*{20}}
\multiput(150, 50)(100,0){2}{\circle*{20}}
\put(150, 50) {\line(1,0){100}}
\put(150, 50) {\line(1,1){ 50}}
\put(250, 50) {\line(-1,1){ 50}}
\put(375,  90) {$\Lambda_Q(nnQ)$}
\put(375,  40) {$\Xi_Q(nsQ)$}
\put(-50,  65) {$(0,1) \equiv {\bf \bar{3}}$}
\multiput(150,-50)(100,0){2}{\circle*{20}}
\put(200,-100) {\circle*{20}}
\put(150, -50) {\line(1,0){100}}
\put(150, -50) {\line(1,-1){ 50}}
\put(250, -50) {\line(-1,-1){ 50}}
\put(375, -60) {$\Xi_{QQ}(QQn)$}
\put(375,-110) {$\Omega_{QQ}(QQs)$}
\put(-50, -85) {$(1,0) \equiv {\bf 3}$}
\end{picture}
\caption{Baryon flavor multiplets: light baryon octet ${\bf 8}$ and decuplet ${\bf 10}$, singly-heavy baryon sextet ${\bf 6}$ and anti-triplet ${\bf \bar{3}}$ and doubly-heavy baryon triplet ${\bf 3}$ configurations. Here $n$ denotes the nonstrange light quarks $n=u$, $d$ and $Q$ the heavy quarks $Q=c$, $b$.}
\label{Baryons}
\end{figure}

\subsection{SU(3) Isoscalar factors}

For the flavor $SU(3)$ we use the phase convention of Baird and Biedenharn \cite{Baird:1963wv}. 
The three light quarks ($u,d,s$) transform according to the fundamental triplet $(p,q)=(1,0)$ and the antiquarks ($\bar{u},-\bar{d},\bar{s}$) according to the anti-triplet $(p,q)=(0,1)$ 
\cite{Ortiz_Pacheco_2019P,Ortiz-Pacheco:2021thesis}. The $SU(3)$ isocalar factors relevant for the strong couplings of heavy baryons can be obtained with the help of Refs.~\cite{10.1063/1.3521562,SU(N)}. The basic process in the elementary emission model in which the meson is emitted by a light quark is shown in Fig.~\ref{qqM}.

\begin{figure}[t]
\centering
\setlength{\unitlength}{0.38pt}
\begin{picture}(550,300)(-50,50)
\thicklines
\multiput(150,300)(100,0){2}{\circle*{20}}
\multiput(100,250)(100,0){3}{\circle*{20}}
\multiput(150,200)(100,0){2}{\circle*{20}}
\put(150,300) {\line(1,0){100}}
\put(100,250) {\line(1,0){200}}
\put(150,200) {\line(1,0){100}}
\put(100,250) {\line(1,1){ 50}}
\put(150,200) {\line(1,1){100}}
\put(250,200) {\line(1,1){ 50}}
\put(300,250) {\line(-1,1){ 50}}
\put(250,200) {\line(-1,1){100}}
\put(150,200) {\line(-1,1){ 50}}
\put(375, 290) {$K(n\bar{s})$}
\put(375, 240) {$\pi, \eta_8 (n\bar{n}, s\bar{s})$}
\put(375, 190) {$\overline{K}(s\bar{n})$}
\put(-50, 240) {$(1,1) \equiv {\bf 8}$}
\put(200,100) {\circle*{20}}
\put(375,  90) {$\eta_1(n\bar{n}, s\bar{s})$}
\put(-50,  90) {$(0,0) \equiv {\bf 1}$}
\end{picture}
\caption{Meson flavor multiplets: light meson octet ${\bf 8}$ and singlet ${\bf 1}$. Here $n$ denotes the nonstrange light quarks $n=u$, $d$ and $Q$ the heavy quarks $Q=c$, $b$.}
\label{Mesons}
\end{figure}

(i) First consider the decay of a singly-heavy baryon into another singly-heavy baryon and a pseudo-scalar meson:
\ba
qqQ \rightarrow qq'Q + q\bar{q}'
\label{strong1}
\ea
The baryon $qqQ$ configuration can either be a light-flavor sextet or anti-triplet, corresponding to the Young tableaux $[f]=[200]$ and $[110]$, respectively. An alternative notation is in terms of $(p,q)=(f_1-f_2,f_2-f_3)$ which gives $(p,q)=(2,0)$ and $(0,1)$. The mesons belong to the octet or singlet configurations with $[f]=[210]$ and $[000]$ or, equivalently, $(p,q)=(1,1)$ and $(0,0)$. 

\noindent
$(20) \rightarrow (20) \otimes (00)$
\ba
\left( \begin{array}{c} \Sigma_{Q} \\ \Xi'_Q \\ \Omega_Q \end{array} \right) \;\rightarrow\; 
\left( \begin{array}{c} \Sigma_Q \eta_1 \\ \Xi'_Q \eta_1 \\ \Omega_Q \eta_1 \end{array} \right) \;=\; 
\left( \begin{array}{c} 1 \\ 1 \\ 1 \end{array} \right)^{1/2}
\ea

\noindent
$(20) \rightarrow (20) \otimes (11)$
\ba
\left( \begin{array}{c} \Sigma_Q \\ \Xi'_Q \\ \Omega_Q \end{array} \right) &\rightarrow& 
\left( \begin{array}{c} \Sigma_Q \pi \quad \Sigma_Q \eta_8 \quad \Xi'_Q K \\ 
\Sigma_Q \overline{K} \quad \Xi'_Q \pi \quad \Xi'_Q \eta_8 \quad \Omega_Q K \\ 
\Xi'_Q \overline{K} \quad \Omega_Q \eta_8 \end{array} \right) 
\nonumber\\
&=& \frac{1}{2\sqrt{10}} \left( \begin{array}{c} 24 \quad -4 \quad 12 \\
18 \quad 9 \quad 1 \quad 12 \\ 24 \quad 16 \end{array} \right)^{1/2}
\ea

\noindent
$(20) \rightarrow (01) \otimes (11)$
\ba
\left( \begin{array}{c} \Sigma_Q \\ \Xi'_Q \\ \Omega_Q \end{array} \right) &\rightarrow& 
\left( \begin{array}{c} \Lambda_Q \pi \quad \Xi_Q K \\ 
\Lambda_Q \overline{K} \quad \Xi_Q \pi \quad \Xi_Q \eta_8 \\ 
\Xi_Q \overline{K} \end{array} \right) 
\nonumber\\
&=& \frac{1}{2\sqrt{2}} \left( \begin{array}{c} 4 \quad -4 \\
2 \quad 3 \quad -3 \\ 8 \end{array} \right)^{1/2}
\ea

\noindent
$(01) \rightarrow (01) \otimes (00)$
\ba
\left( \begin{array}{c} \Lambda_Q \\ \Xi_Q \end{array} \right) \;\rightarrow\; 
\left( \begin{array}{c} \Lambda_Q \eta_1 \\ \Xi_Q \eta_1 \end{array} \right) \;=\; 
\left( \begin{array}{c} 1 \\ 1 \end{array} \right)^{1/2}
\ea

\noindent
$(01) \rightarrow (01) \otimes (11)$
\ba
\left( \begin{array}{c} \Lambda_Q \\ \Xi_Q \end{array} \right) &\rightarrow& 
\left( \begin{array}{c} \Lambda_Q \eta_8 \quad \Xi_Q K \\ 
\Lambda_Q \overline{K} \quad \Xi_Q \pi \quad \Xi_Q \eta_8 \end{array} \right) 
\nonumber\\
&=& \frac{1}{4} \left( \begin{array}{c} 4 \quad -12 \\ 6 \quad -9 \quad -1 \end{array} \right)^{1/2}
\ea

\noindent
$(01) \rightarrow (20) \otimes (11)$
\ba
\left( \begin{array}{c} \Lambda_Q \\ \Xi_Q \end{array} \right) &\rightarrow& 
\left( \begin{array}{c} \Sigma_Q \pi \quad \Xi'_Q K \\ 
\Sigma_Q \overline{K} \quad \Xi'_Q \pi \quad \Xi'_Q \eta_8 \quad \Omega_Q K \end{array} \right) 
\nonumber\\
&=& \frac{1}{4} \left( \begin{array}{c} -12 \quad -4 \\ -6 \quad 3 \quad -3 \quad 4 \end{array} \right)^{1/2}
\ea

(ii) Subsequently, we analyze the decay of a doubly-heavy baryon into another doubly-heavy baryon and a light pseudo-scalar meson  
\ba
QQq \rightarrow QQq' + q\bar{q}'
\label{strong3}
\ea
are characterized by the following couplings.

\noindent
$(10) \rightarrow (10) \otimes (00)$
\ba
\left( {\begin{array}{c} \Omega_{QQ} \\ \Xi_{QQ} \end{array} } \right) \;\rightarrow\; 
\left( \begin{array}{c} \Omega_{QQ}\eta_1 \\ \Xi_{QQ}\eta_1 \end{array} \right)
\;=\; \left( \begin{array}{c} 1 \\ 1 \end{array} \right)^{1/2}
\ea

\noindent
$(10) \rightarrow (10) \otimes (11)$
\ba
\left( {\begin{array}{c} \Omega_{QQ} \\ \Xi_{QQ} \end{array} } \right) &\rightarrow& 
\left( \begin{array}{c} \Omega_{QQ}\eta_8 \quad	\Xi_{QQ}\overline{K} \\
\Omega_{QQ}K \quad \Xi_{QQ}\eta_8 \quad \Xi_{QQ}\pi \end{array} \right)
\nonumber\\
&=& \frac{1}{4}\left( \begin{array}{c} 4 \quad 12 \\
6 \quad -1 \quad 9 \end{array} \right)^{1/2}
\ea

The corresponding $SU(3)$ reduced matrix elements are given in Table~\ref{su3rme}. For the decays of Eqs.~(\ref{strong1}) the contribution arises from the first two quarks ($j=1,2$) and for the decays in Eqs.~(\ref{strong3}) from the third quark ($j=3$).

\begin{table}[t]
\centering
\caption{$SU(3)$ reduced matrix elements for singly-heavy baryons (top) and doubly-heavy baryons (bottom).}
\label{su3rme}
\begingroup
\footnotesize
\setlength{\tabcolsep}{1.8pt}
\renewcommand{\arraystretch}{1.00}
\begin{tabular}{cccccc}
\noalign{\smallskip}
\toprule
& & & \multicolumn{3}{c}{$\langle (p',q') \parallel T^{(p'',q'')}_j \parallel (p,q) \rangle$} \\
$(p,q)$ & $(p',q')$ & $(p'',q'')$ & $j=1$ & $j=2$ & $j=3$ \\ 
\noalign{\smallskip}
\hline
\noalign{\smallskip}
$(2,0)$ & $(2,0)$ & $(0,0)$ & $\frac{\sqrt{2}}{\sqrt{3}}$ & $\frac{\sqrt{2}}{\sqrt{3}}$ & $0$ \\
\noalign{\smallskip}
$(2,0)$ & $(2,0)$ & $(1,1)$ & $\frac{\sqrt{10}}{\sqrt{3}}$ & $\frac{\sqrt{10}}{\sqrt{3}}$ & $0$ \\
\noalign{\smallskip}
$(2,0)$ & $(0,1)$ & $(1,1)$ & $\sqrt{2}$ & $-\sqrt{2}$ & $0$ \\
\noalign{\smallskip}
$(0,1)$ & $(0,1)$ & $(0,0)$ & $\frac{\sqrt{2}}{\sqrt{3}}$ & $\frac{\sqrt{2}}{\sqrt{3}}$ & $0$ \\
\noalign{\smallskip}
$(0,1)$ & $(0,1)$ & $(1,1)$ & $-\frac{2}{\sqrt{3}}$ & $-\frac{2}{\sqrt{3}}$ & $0$ \\
\noalign{\smallskip}
$(0,1)$ & $(2,0)$ & $(1,1)$ & $2$ & $-2$ & $0$ \\
\noalign{\smallskip}
\hline
\noalign{\smallskip}
$(1,0)$ & $(1,0)$ & $(0,0)$ & $0$ & $0$ & $\frac{\sqrt{2}}{\sqrt{3}}$ \\
\noalign{\smallskip}
$(1,0)$ & $(1,0)$ & $(1,1)$ & $0$ & $0$ & $\frac{4}{\sqrt{3}}$ \\
\noalign{\smallskip}
\botrule
\end{tabular}
\endgroup
\end{table}

Finally, the matrix elements of the spin operators are presented in Table~\ref{spinme}.
\begin{table}[t]
\centering
\caption{Spin matrix elements.}
\label{spinme}
\begin{tabular}{rcccc}
\toprule
\noalign{\smallskip}
& $\nu$ & $s_{1-}$ & $s_{2-}$ & $s_{3-}$ \\
\noalign{\smallskip}
\hline
\noalign{\smallskip}
$\chi_{S,3/2} \rightarrow \chi_{\rho,1/2}$ & $\frac{1}{2}$ & $-\frac{1}{\sqrt{2}}$ & $\frac{1}{\sqrt{2}}$ & $0$ \\
\noalign{\smallskip}
$\rightarrow \chi_{\lambda,1/2}$ & $\frac{1}{2}$ & $-\frac{1}{\sqrt{6}}$ & $-\frac{1}{\sqrt{6}}$ & $\frac{\sqrt{2}}{\sqrt{3}}$ \\
\noalign{\smallskip}
$\rightarrow \chi_{S,1/2}$ & $\frac{1}{2}$ & $\frac{1}{\sqrt{3}}$ & $\frac{1}{\sqrt{3}}$ & $\frac{1}{\sqrt{3}}$ \\
\noalign{\smallskip}
\hline
\noalign{\smallskip}
& $\nu$ & $s_{10}$ & $s_{20}$ & $s_{30}$ \\
\noalign{\smallskip}
\hline
\noalign{\smallskip}
$\chi_{\rho,1/2} \rightarrow \chi_{\rho,1/2}$ & $\frac{1}{2}$ & $0$ & $0$ & $\frac{1}{2}$ \\
\noalign{\smallskip}
$\rightarrow \chi_{\lambda,1/2}$ & $\frac{1}{2}$ & $-\frac{1}{2\sqrt{3}}$ & $\frac{1}{2\sqrt{3}}$ & $0$ \\
\noalign{\smallskip}
$\rightarrow \chi_{S,1/2}$ & $\frac{1}{2}$ & $\frac{1}{\sqrt{6}}$ & $-\frac{1}{\sqrt{6}}$ & $0$ \\
\noalign{\smallskip}
$\chi_{\lambda,1/2} \rightarrow \chi_{\rho,1/2}$ & $\frac{1}{2}$ & $-\frac{1}{2\sqrt{3}}$ & $\frac{1}{2\sqrt{3}}$ & $0$ \\
\noalign{\smallskip}
$\rightarrow \chi_{\lambda,1/2}$ & $\frac{1}{2}$ & $\frac{1}{3}$ & $\frac{1}{3}$ & $-\frac{1}{6}$ \\
\noalign{\smallskip}
$\rightarrow \chi_{S,1/2}$ & $\frac{1}{2}$ & $\frac{1}{3\sqrt{2}}$ & $\frac{1}{3\sqrt{2}}$ & $-\frac{\sqrt{2}}{3}$ \\
\noalign{\smallskip}
$\chi_{S,1/2} \rightarrow \chi_{\rho,1/2}$ & $\frac{1}{2}$ & $\frac{1}{\sqrt{6}}$ & $-\frac{1}{\sqrt{6}}$ & $0$ \\
\noalign{\smallskip}
$\rightarrow \chi_{\lambda,1/2}$ & $\frac{1}{2}$ & $\frac{1}{3\sqrt{2}}$ & $\frac{1}{3\sqrt{2}}$ & $-\frac{\sqrt{2}}{3}$ \\
\noalign{\smallskip}
$\rightarrow \chi_{S,1/2}$ & $\frac{1}{2}$ & $\frac{1}{6}$ & $\frac{1}{6}$ & $\frac{1}{6}$ \\
\noalign{\smallskip}
$\chi_{S,3/2} \rightarrow \chi_{S,3/2}$ & $\frac{3}{2}$ & $\frac{1}{2}$ & $\frac{1}{2}$ & $\frac{1}{2}$ \\
\noalign{\smallskip}
\hline
\noalign{\smallskip}
& $\nu$ & $s_{1+}$ & $s_{2+}$ & $s_{3+}$ \\
\noalign{\smallskip}
\hline
\noalign{\smallskip}
$\chi_{\rho,-1/2} \rightarrow \chi_{\rho,1/2}$ & $\frac{1}{2}$ & $0$ & $0$ & $1$ \\
\noalign{\smallskip}
$\rightarrow \chi_{\lambda,1/2}$ & $\frac{1}{2}$ & $-\frac{1}{\sqrt{3}}$ & $\frac{1}{\sqrt{3}}$ & $0$ \\
\noalign{\smallskip}
$\rightarrow \chi_{S,1/2}$ & $\frac{1}{2}$ & $-\frac{1}{\sqrt{6}}$ & $\frac{1}{\sqrt{6}}$ & $0$ \\
\noalign{\smallskip}
$\chi_{\lambda,-1/2} \rightarrow \chi_{\rho,1/2}$ & $\frac{1}{2}$ & $-\frac{1}{\sqrt{3}}$ & $\frac{1}{\sqrt{3}}$ & $0$ \\
\noalign{\smallskip}
$\rightarrow \chi_{\lambda,1/2}$ & $\frac{1}{2}$ & $\frac{2}{3}$ & $\frac{2}{3}$ & $-\frac{1}{3}$ \\
\noalign{\smallskip}
$\rightarrow \chi_{S,1/2}$ & $\frac{1}{2}$ & $-\frac{1}{3\sqrt{2}}$ & $-\frac{1}{3\sqrt{2}}$ & $\frac{\sqrt{2}}{3}$ \\
\noalign{\smallskip}
$\chi_{S,-1/2} \rightarrow \chi_{\rho,1/2}$ & $\frac{1}{2}$ & $\frac{1}{\sqrt{6}}$ & $-\frac{1}{\sqrt{6}}$ & $0$ \\
\noalign{\smallskip}
$\rightarrow \chi_{\lambda,1/2}$ & $\frac{1}{2}$ & $\frac{1}{3\sqrt{2}}$ & $\frac{1}{3\sqrt{2}}$ & $-\frac{\sqrt{2}}{3}$ \\
\noalign{\smallskip}
$\rightarrow \chi_{S,1/2}$ & $\frac{1}{2}$ & $\frac{2}{3}$ & $\frac{2}{3}$ & $\frac{2}{3}$ \\
\noalign{\smallskip}
$\chi_{\rho,1/2} \rightarrow \chi_{S,3/2}$ & $\frac{3}{2}$ & $-\frac{1}{\sqrt{2}}$ & $\frac{1}{\sqrt{2}}$ & $0$ \\
\noalign{\smallskip}
$\chi_{\lambda,1/2} \rightarrow \chi_{S,3/2}$ & $\frac{3}{2}$ & $-\frac{1}{\sqrt{6}}$ & $-\frac{1}{\sqrt{6}}$ 
& $\frac{\sqrt{2}}{\sqrt{3}}$ \\
\noalign{\smallskip}
$\chi_{S,1/2} \rightarrow \chi_{S,3/2}$ & $\frac{3}{2}$ & $\frac{1}{\sqrt{3}}$ & $\frac{1}{\sqrt{3}}$ & $\frac{1}{\sqrt{3}}$ \\
\noalign{\smallskip}
\botrule
\end{tabular}
\end{table}

\section{Strong decay widths}
\label{widths}

The width of a baryon decaying into a baryon and a (pseudoscalar) meson is given by
\ba
\Gamma(B \rightarrow B'+ M)=2\pi \rho \frac{2}{2J+1}\sum_{\nu>0}|\mathcal{A}_\nu(k)|^2 ~. \label{width_formula} 
\ea
Here $\rho$ denotes the phase-space factor calculated in the rest frame of the initial baryon 
\ba
\rho=4\pi\frac{E_{B'}E_{M}k}{m_{B}} ~,\label{phase_space}
\ea 
and the magnitude square of the three-momentum from the emitted meson is
\ba
k^2=-m_M^2+\frac{(m_{B}^2-m_{B'}^2+m_M^2)^2}{4m_{B}^2} ~.
\ea
The energies of the final baryon and meson are given by $E_{B'}=\sqrt{m_{B'}^2+k^2}$ and $E_{M}=\sqrt{m_{M}^2+k^2}$, respectively. 

The physical $\eta$ and $\eta'$ mesons are mixtures of the $\eta$ meson of the octet and the singlet, $\eta_8$ and $\eta_1$, 
\ba
\eta  &=& \eta_8 \cos \theta_P - \eta_1 \sin \theta_P ~, 
\nonumber\\
\eta' &=& \eta_8 \sin \theta_P + \eta_1 \cos \theta_P ~,
\ea
where the mixing angle is given by $\theta_P = -23^\circ$ \cite{PhysRevD.55.2862, PhysRevD.50.2048}. As a consequence, the flavor operators for these mesons are mixed in the similar way 
\ba
X^\eta &=& X^{\eta_8} cos \theta_P-X^{\eta_1} \sin \theta_P ~,
\nonumber\\
X^{\eta'} &=& X^{\eta_8} \sin \theta_P+X^{\eta_1} \cos \theta_P ~.
\ea

The coupling constants $g$ and $h$ were determined in a global fit to the 52 experimental baryon decay widths (32 singly charm baryons and 20 singly bottom baryons) presented in Tables~\ref{SigmaQ}-\ref{OmegaQ} \cite{PDG26} with the exception of $\Lambda_c(2765)^+$ for which no experimental uncertainty has been reported. Following our previous study on the masses of heavy charm and bottom baryons \cite{EOP2023}, we construct and minimize a likelihood function for the strong decay widths. This approach incorporates the measured decay widths $\Gamma_{\text{exp}}$ and their associated uncertainties to obtain the optimal values of $g$ and $h$ to be $g=3.32$ GeV$^{-1}$ and $h=-0.16$ GeV$^{-1}$. These are the values used to calculate the decay widths of all heavy baryons considered in the present work. The values of the quark masses are given by $m_u=m_d=291.53$ MeV, $m_s=461.24$ MeV, $m_c=1606.80$ MeV and $m_b=4944.34$ MeV (taken from Table~I of Ref.~\cite{EOP2023}).

Tables~\ref{SigmaQ}-\ref{OmegaQ} show the partial and total strong decay widths calculated in the elementary meson emission model. A dash $-$ indicates that the decay is kinematically forbidden by energy conservation, a $0$ that the decay vanishes identically because of a selection rule, and $0.0$ that the value of the decay width is zero at the current precision. The baryon masses are taken from Ref.~\cite{EOP2023}. The EEM results are compared with those of two other calculations, the $^3P_0$ model of Refs.~\cite{PhysRevD.107.034031,garciatecocoatzi2023decay} and the chiral quark model ($\chi$QM) of Refs.~\cite{Wang2017,WangOme}, both of which present a comprehensive study of the $S$- and $P$-wave heavy baryons of the sextet, $\Sigma_Q$, $\Xi'_Q$ and $\Omega_Q$, and the anti-triplet, $\Lambda_Q$ and $\Xi_Q$.

\subsection{Selection rules}

There are several decays of $P$-wave heavy baryons that are forbidden by selection rules. 
For example, if we consider the decays of sextet baryons into an anti-triplet baryon and an octet meson, $(2,0) \rightarrow (0,1) \otimes (1,1)$, the Table~\ref{su3rme} shows that in the flavor sector only the light baryons with $j=1,2$ contribute. For the $^{2}\rho$ configuration of the initial baryon the spin wave function is given by $\chi_\rho$, just as for the final baryon \cite{EOP2023}. According to Table~\ref{spinme}, the only non-vanishing contribution in the spin sector arises from the heavy quark with $j=3$. As a result, the strong decays of the $^{2}\rho$ configuration of sextet baryons to an anti-triplet baryon and an octet meson is forbidden by the structure of the spin-flavor wave functions.

Another example are the baryon decays from the anti-triplet into another baryon from the anti-triplet and meson from the octet, $(0,1) \rightarrow (0,1) \otimes (1,1)$. Also in this case, the only contribution from the flavor part comes from the light baryons with $j=1,2$. For the $^{2}\lambda$ configuration of the initial baryon the spin wave function is given by $\chi_\rho$, just as for the final baryon which means that in spin space the only contribution comes from the heavy quark with $j=3$. As a result, also this class of strong decays is forbidden by the structure of spin-flavor wave function. 

Finally, the strong decay of the $^{2}\rho$ configuration of a flavor triplet double heavy baryon into another flavor triplet heavy baryon and an octet meson, $(1,0) \rightarrow (1,0) \otimes (1,1)$ is forbidden by the structure of the spin-flavor wave functions. In the flavor sector the only contribution comes from the heavy quark with $j=3$, whereas the spin contribution arises from the first two heavy quarks.

\subsection{$\Sigma_Q$ baryons}

Table~\ref{SigmaQ} shows the partial and total decay widths of the $S$- and $P$-wave $\Sigma_c$ baryons calculated in the EEM. A comparison with the available experimental data shows a reasonable agreement especially if one takes into account the large experimental uncertainties for the decay widths of the charged  $\Sigma_c(2800)$ baryons. The decays of the $\Sigma_c(2455)$ and $\Sigma_c(2520)$ baryons are dominated by the $\Lambda_c\pi$ channel with approximately 100 \% of the branching fraction \cite{PhysRevD.89.091102,PDG26}, consistent with an interpretation as ground state $S$-wave baryons, $^2(\Sigma)_{1/2}$ and $^4(\Sigma)_{3/2}$. 

The $\Sigma_c(2800)$ resonance was assigned as the $P$-wave state $^2\lambda(\Sigma_{c})_{{1/2}^-}$ \cite{EOP2023}. The decay width was reported by the Belle Collaboration \cite{PhysRevLett.94.122002} and later confirmed by BaBar \cite{PhysRevD.78.112003}, albeit with a large experimental uncertainty. In the EEM, the calculated width arises mostly from $\Sigma_c\pi$ and $\Lambda_c\pi$ channels. 

The remaining $P$-wave charm baryons decay widths are calculated to be in the range between 12 and 54 MeV, for which there are no experimental information available so far. The partial decay widths of the $^{2S+1}\rho(\Sigma_c)_J$ configurations to the $\Lambda_c \pi$, $\Xi_c \pi$ and $ND$ channel are forbidden. As a consequence, the calculated decay widths for these baryons arise exclusively from the $\Sigma_c \pi$ and $\Sigma_c^\ast \pi$ channels and are calculated to be of the order of 40-50 MeV. 

\begin{table*}[t]
\caption{Partial and total decay widths of sextet $\Sigma_Q$ baryons in MeV.}
\label{SigmaQ}
\centering
\begingroup
\setlength{\tabcolsep}{1.8pt}
\renewcommand{\arraystretch}{1.02}
\begin{adjustbox}{max width=\textwidth,center}
\begin{tabular}{ccrrrrrrrcl}
\noalign{\smallskip}
\toprule %
& \multicolumn{6}{c}{Present work}  && \\ 
\noalign{\smallskip}
\cmidrule(lr){2-7} 
\noalign{\smallskip}
State & $M^{th}(\text{MeV})$ & $\Sigma_c\pi$ & $\Sigma^\ast_c\pi$ & $\Lambda_c\pi$ & $\Xi_cK$ & $\Gamma_{\text{EEM}}$ & $^3P_0$ \cite{PhysRevD.107.034031} & $\chi$QM \cite{Wang2017} & $\Gamma_{\text{exp}}$ \cite{PDG26} & Baryon \\
\noalign{\smallskip}
\hline
\noalign{\smallskip}
$^2(\Sigma_{c})_{1/2}$  &  $2455 \pm 2$    &   -- &  1.8 & -- &  -- &  1.8 &   1.7 
& &  $1.89^{+0.09}_{-0.18}$ & $\Sigma_{c}(2455)^{++}$ \\
&  & -- &   -- &  1.8 & -- &  1.8 & 1.7 
& & $2.3 \pm 0.4$ & $\Sigma_{c}(2455)^{+}$ \\
&  & -- &   -- &  1.8 & -- &  1.8 & 1.7 
& & $1.83 ^{+0.11}_{-0.19}$ & $\Sigma_{c}(2455)^{0}$ \\
$^4(\Sigma_{c})_{3/2}$ & $2521 \pm 2$       &   -- & 11.6 & -- &  -- & 11.6 &  14.9 
& & $14.78^{+0.30}_{-0.40}$ & $\Sigma_{c}(2520)^{++}$ \\
&  &  -- &   -- & 11.6 & -- & 11.6 & 14.9 
& &  $17.2^{+4.0}_{-2.2}$  & $\Sigma_{c}(2520)^{+}$ \\
& &  -- &   -- & 11.6 &  -- & 11.6 & 14.9 
& &  $15.3^{+0.4}_{-0.5}$  & $\Sigma_{c}(2520)^{0}$ \\
$^2\lambda(\Sigma_{c})_{1/2}$ & $2890 \pm 2$ & 3.5 &  0.8 &  8.4 &  -- & 12.7 &  30.4 
& 22.65 & $75^{+22}_{-17}$ & $\Sigma_{c}(2800)^{++}$ \\
&  &3.5 &  0.8 &  8.4 & -- & 12.7 & 30.4 
& 22.65 & $62 ^{+60}_{-40}$ & $\Sigma_{c}(2800)^{+}$ \\
&  &3.5 &  0.8 &  8.4 & -- & 12.7 & 30.4 
& 22.65 & $72 ^{+22}_{-15}$ & $\Sigma_{c}(2800)^{0}$ \\
$^2\lambda(\Sigma_{c})_{3/2}$ & $2922 \pm 2$& 7.9 &  1.2 & 12.9 & -- & 22.0 
& 182.7 & 36.50 && \\
$^4\lambda(\Sigma_{c})_{1/2}$ & $2923 \pm 2$& 2.9 &  0.8 & 21.2 & -- & 25.6 
&  61.2 & 17.63 && \\
$^4\lambda(\Sigma_{c})_{3/2}$ & $2956 \pm 2$& 0.6 &  8.7 &  3.1 & -- & 12.4 
& 165.1 & 24.69 && \\
$^4\lambda(\Sigma_{c})_{5/2}$ &  $3011 \pm 2$&5.9 & 11.0 & 24.6 & -- & 41.7 
& 236.5 & 33.22 && \\
$^2\rho(\Sigma_{c})_{1/2}$    & $3029 \pm 2$&16.5 & 24.4 &    0 & -- & 40.9 
& 124.6 & && \\
$^2\rho(\Sigma_{c})_{3/2}$    & $3062 \pm 2$ &25.2 & 29.1 &    0 & 0 & 54.2 
& 125.0 & && \\
\noalign{\smallskip}
\hline
\noalign{\smallskip}
& \multicolumn{6}{c}{Present work}  && \\ 
\noalign{\smallskip}
\cmidrule(lr){2-7} 
\noalign{\smallskip}
State & $M^{th}(\text{MeV})$  & $\Sigma_b\pi$ & $\Sigma^\ast_b\pi$ & $\Lambda_b\pi$ & $\Xi_cK$ & $\Gamma_{\text{EEM}}$ & $^3P_0$ \cite{garciatecocoatzi2023decay} & $\chi$QM \cite{Wang2017} & $\Gamma_{\text{exp}}$ \cite{PDG26,PhysRevLett.122.012001} & Baryon \\
\noalign{\smallskip}
\hline
\noalign{\smallskip}
$^2(\Sigma_{b})_{1/2}$        &  $5810 \pm 3$& -- &   -- &  4.6 & -- &  4.6 &   3.9 
& &  $5.0 \pm 0.5$ & $\Sigma_{b}^+$ \\
&  & -- &   -- &  4.6 & -- &  4.6 &   3.9 
& &  $5.3 \pm 0.5$ & $\Sigma_{b}^-$ \\
$^4(\Sigma_{b})_{3/2}$  & $5832.5 \pm 0.2$&  -- &   -- &  8.3 & -- &  8.3 &  10.0 
& & $9.4 \pm 0.5$   & $\Sigma^{*+}_{b}$ \\      
&  & -- &   -- &  8.3 & -- &  8.3 &  10.0 
& & $10.4 \pm 0.8$   & $\Sigma^{*-}_{b}$ \\
$^2\lambda(\Sigma_{b})_{1/2}$ & $6096.9 \pm 1.2$ & 2.8 &  1.4 & 12.2 & -- & 16.3 
&  23.5 & 22.66 & $31.0 \pm 5.5$ & $\Sigma_{b}(6097)^+$ \\
&   &2.8 &  1.4 & 12.2 & -- & 16.3 
&  23.5 & 22.66 & $28.9 \pm 4.3$ & $\Sigma_{b}(6097)^-$ \\
$^2\lambda(\Sigma_{b})_{3/2}$ &  $6114 \pm 3$  &4.7 &  1.3 & 15.7 & -- & 21.7 
&  83.9 & 39.29 && \\
$^4\lambda(\Sigma_{b})_{1/2}$ & $6123 \pm 3$& 1.9 &  1.0 & 27.3 & -- & 30.1 
&  13.2 & 14.21 && \\
$^4\lambda(\Sigma_{b})_{3/2}$ &  $6130 \pm 3$&0.3 &  7.1 &  3.5 & -- & 10.9 
&  56.9 & 26.29 && \\
$^4\lambda(\Sigma_{b})_{5/2}$ &  $6141 \pm 3$&2.2 &  5.6 & 22.2 & -- & 30.0 
&  95.9 & 38.34 && \\
$^2\rho(\Sigma_{b})_{1/2}$    &$6300 \pm 3$& 16.7 & 35.4 &    0 &  0 & 52.1 
& 134.2 & && \\
$^2\rho(\Sigma_{b})_{3/2}$    & $6307 \pm 3$&21.5 & 33.7 &    0 &  0 & 55.1 
& 129.3 & && \\
\noalign{\smallskip}
\botrule
\end{tabular}
\end{adjustbox}
\endgroup
\end{table*}

Concerning the bottom sector, $\Sigma^\pm_b$ and $\Sigma^{*\pm}_b$ baryons were observed in 2019 by the LHCb Collaboration in the $\Lambda^0 \pi^+$ and $\Lambda^0 \pi^-$ systems which were interpreted as ground state $S$-wave baryons $^2(\Sigma_b)_{J=1/2}$ and $^2(\Sigma^\ast_b)_{J=3/2}$, respectively \cite{PhysRevLett.122.012001}. The decay widths of these resonances are due to the $\Lambda^0_b \pi^\pm$ channel, see Table~\ref{SigmaQ}. In the same experiment, the resonances $\Sigma_b(6097)^\pm$ were interpreted as $P$-wave baryons with a relative angular momentum $L=1$ between $\Lambda^0_b$ and $\pi^\pm$. In our study these baryons are assigned as one $\lambda$-excitation with  $^2\lambda(\Sigma_b)_{1/2}$.

The results for the $\Sigma_b$ baryons show a similar pattern as those for $\Sigma_c$, with the only difference being the mass of the heavy quark. Once again, we find a good overall agreement with the available experimental data, thereby  confirming the assignments made previously  \cite{EOP2023}.

For the $\Sigma_c$ and $\Sigma_b$ sectors one expects the existence of two ground state $S$-wave configurations with $J^P=1/2^+$ and $3/2^+$, both of which have been observed as well as seven $P$-wave baryons, five belonging to the $\lambda$ excitation and two to the $\rho$ excitation. So far, only one of the $P$-wave baryons has been observed, $\Sigma_c(2800)$ and $\Sigma_b(6097)$. 

\subsection{$\Lambda_Q$ baryons}

\begin{table*}[t]
\caption{Partial and total decay widths of anti-triplet $\Lambda_Q$ baryons in MeV.}
\label{LambdaQ}
\centering
\begingroup
\setlength{\tabcolsep}{2.0pt}
\renewcommand{\arraystretch}{1.02}
\begin{adjustbox}{max width=\textwidth,center}
\begin{tabular}{ccrrrrrcl}
\noalign{\smallskip}
\toprule
& \multicolumn{5}{c}{Present work}  & \\ 
\noalign{\smallskip}
\cmidrule(lr){2-6} 
\noalign{\smallskip}
State &  $M^{th}(\text{MeV})$&$\Sigma_c\pi$ & $\Sigma^\ast_c\pi$ & $\Lambda_c \eta$ & $\Gamma_{\text{EEM}}$ 
& $^3P_0$ \cite{PhysRevD.107.034031} & $\Gamma_{\text{exp}}$ \cite{PDG26} & Baryon \\
\noalign{\smallskip}
\hline
\noalign{\smallskip}
$^2(\Lambda_{c})_{1/2}$       & $2281 \pm 2$ &   -- &   -- &  -- &  -- & 
& & $\Lambda_{c}^+$ \\
$^2\lambda(\Lambda_{c})_{1/2}$ & $2616 \pm 2$ &  0.0 &   -- &  -- &  0.0 &   1.4 & $2.59 \pm 0.56$ & $\Lambda_{c}(2595)^+$ \\
$^2\lambda(\Lambda_{c})_{3/2}$ & $2648 \pm 2$ &  0.2 &   -- &  -- &  0.2 &   9.8 & $<0.52$       & $\Lambda_{c}(2625)^+$ \\
$^2\rho(\Lambda_{c})_{1/2}$   & $2788 \pm 2$  &  4.8 &  1.0 &  -- &  5.8 &  56.9 & $50$ & $\Lambda_{c}(2765)^+$ \\
$^2\rho(\Lambda_{c})_{3/2}$   & $2821 \pm 2$  & 10.7 &  1.7 &  -- & 12.4 &  92.7 & & \\
$^4\rho(\Lambda_{c})_{1/2}$   & $2821 \pm 2$  &  4.0 &  1.0 &  -- &  5.0 &  32.5 & & \\
$^4\rho(\Lambda_{c})_{3/2}$   & $2854 \pm 2$  &  0.8 & 11.7 & 0.0 & 12.5 & 121.9 & & \\
$^4\rho(\Lambda_{c})_{5/2}$   & $2909 \pm 2$  &  8.1 & 15.0 & 0.4 & 23.5 & 108.1 & $20^{+6}_{-5}$    & $\Lambda_{c}(2940)^+$  \\
\noalign{\smallskip}
\hline
\noalign{\smallskip}
& \multicolumn{5}{c}{Present work}  & \\ 
\noalign{\smallskip}
\cmidrule(lr){2-6} 
\noalign{\smallskip}
State& $M^{th}(\text{MeV})$ & $\Sigma_b\pi$ & $\Sigma^\ast_b\pi$ & $\Lambda_b \eta$ & $\Gamma_{\text{EEM}}$
& $^3P_0$ \cite{garciatecocoatzi2023decay} & $\Gamma_{\text{exp}}$ \cite{PDG26} & Baryon \\
\noalign{\smallskip}
\hline
\noalign{\smallskip}
$^2(\Lambda_{b})_{1/2}$       & $5615 \pm 2$  &  -- &  -- & -- &  -- & & & $\Lambda_{b}^0$ \\
$^2\lambda(\Lambda_{b})_{1/2}$ &  $5913 \pm 2$&  -- &  -- & -- &  -- & & $<0.25$ & $\Lambda_{b}(5912)^0$ \\
$^2\lambda(\Lambda_{b})_{3/2}$ & $5919 \pm 2$ &  -- &  -- & -- &  -- & & $<0.19$ & $\Lambda_{b}(5920)^0$ \\
$^2\rho(\Lambda_{b})_{1/2}$    & $6106 \pm 2$  & 2.7 & 1.4 & -- & 4.1 &  66.8 & & \\
$^2\rho(\Lambda_{b})_{3/2}$   &  $6112 \pm 2$  & 4.7 & 1.3 & -- & 5.9 &  84.7 & & \\
$^4\rho(\Lambda_{b})_{1/2}$   & $6121 \pm 3$   & 1.9 & 0.9 & -- & 2.8 &  35.5 & & \\
$^4\rho(\Lambda_{b})_{3/2}$   & $6128 \pm 3$  & 0.3 & 7.0 & -- & 7.3 & 127.8 & & \\
$^4\rho(\Lambda_{b})_{5/2}$   & $6138 \pm 3$   & 2.2 & 5.5 & -- & 7.7 &  74.3 & & \\
\noalign{\smallskip}
\botrule
\end{tabular}
\end{adjustbox}
\endgroup
\end{table*}

The $\Lambda_Q$ states belong to the flavor anti-triplet for which the quark model predicts the existence of one ground state $S$-wave configuration with $J^P=1/2^+$ and seven $P$-wave baryons, two belonging to the $\lambda$ excitation and five to the $\rho$ excitation. The ground state $S$-wave baryons have been observed, $\Lambda_c^+$ and $\Lambda_b^0$, as well as the two $P$-wave states of the $^2\lambda(\Lambda_Q)_{J=1/2,3/2}$ configuration: $\Lambda_c(2595)^+$, $\Lambda_c(2625)^+$ \cite{PhysRevD.84.012003,JHEP05(2017)030,PhysRevLett.98.012001,PhysRevLett.98.262001} and $\Lambda_b(5912)^0$, $\Lambda_b(5920)^0$ \cite{PhysRevLett.109.172003,lamb,PhysRevLett.123.152001,PhysRevD.88.071101,PLB803135345}, respectively. So far, only two excited $\Lambda_c$ baryons have been assigned as belonging to the $\rho$ excitations in the charm sector, $\Lambda_c(2765)^+$ and $\Lambda_c(2940)^+$. 

These assignments are confirmed by the analysis of the EEM strong decay widths presented in Table~\ref{LambdaQ}. Note that PDG does not report an uncertainty for the width of $\Lambda_c(2765)^+$ baryon \cite{PDG26}. All calculated widths in the EEM are relatively small $\lesssim 12.5$ MeV with the exception of that of the $^4\rho(\Lambda_c)_{5/2}$ configuration. 

The calculated widths are in reasonable agreement with the experimental data with the exception of $\Lambda_c(2765)^+$ whose calculated value of 5.8 MeV is much smaller than the experimental one of 50 MeV. One has to take into account that the experimental status of this resonance is very uncertain. It was assigned only a single star and was omitted from the Particle Data Group Summary Table \cite{PDG26}. Moreover, no uncertainty has been reported by PDG for its decay width. In Ref.~\cite{PhysRevLett.98.262001}, the Belle Collaboration determined the width of this resonance to be $73\pm 5$ MeV.

Interestingly, there are various interpretations of this state in the literature. In particular, Refs.~\cite{PhysRevD.103.094003,PhysRevD.105.074036,PhysRevD.101.094023} propose that $\Lambda_{c}(2765)^+$ could be a candidate for a ``Roper-like'' baryon due to its substantial excitation energy of approximately 500 MeV above the ground state, which is consistent with a possible spin-parity assignment of $J^P = \frac{1}{2}^+$. Under this assumption, Ref.~\cite{PhysRevD.103.094003} estimates an upper limit to the decay width of 49.0 MeV, which is close to the experimental value, while the lower limit, 10.8 MeV, is closer to our result. Such an interpretation would correspond to a $2S$-wave baryon with two excitation quanta in the harmonic oscillator classification. The $2S$-wave scenario is beyond the scope of the present investigation and may be considered in subsequent studies. 
 
The Particle Data Group reports two other $\Lambda_c$ baryons, $\Lambda_c(2860)^+$ and $\Lambda_c(2880)^+$, which were tentatively assigned with angular momentum and parity $J^P=3/2^+$ and $5/2^+$, respectively. In the harmonic oscillator quark model this would correspond to two quanta of excitation. In the present analysis we limit ourselves to ground state $1S$-wave and excited state $1P$-wave baryons. The study of higher excited state will be presented in a future publication. 

\subsection{$\Xi_Q$ and $\Xi'_Q$ baryons}

The quark content of $\Xi_Q$ and $\Xi'_Q$ baryons is $nsQ$, {\it i.e.} one heavy quark $Q$, one strange quark $s$ and one non-strange quark $n=u,d$. The $\Xi'_Q$ baryons belong to the flavor sextet and the $\Xi_Q$ baryons to the flavor anti-triplet. All of these states are isospin doublets. Table~\ref{XiQ} shows that 11 of these isospin doublet have been observed, 7 in the charm sector and 4 in the bottom sector. 
We follow the assignments of the strange-charm $\Xi_Q$ and $\Xi'_Q$ baryons from Ref.~\cite{PDG26} in the harmonic oscillator quark models of Refs.~\cite{PhysRevD.105.074029,EOP2023}. All ground state $1S$-wave baryons have been identified.  The situation for the $1P$-wave baryons is very different. In the charm sector ($Q=c$) candidates for half of the in total 14 $P$-wave states have been assigned, and in the bottom sector only 3 out of 14. 

Recently, the spin and parity of the $\Xi_c(3055)^{+,0}$ baryons were determined experimentally by the LHCb Collaboration to be \(J^P=\tfrac{3}{2}^+\), supporting its assignment as a $D$-wave $\lambda$-mode excitation of the $\Xi_c$ anti-triplet~\cite{PhysRevLett.134.081901} 
rather than as a $P$-wave $\lambda$-mode \cite{EOP2023}.

The LHCb Collaboration reported the discovery of three new resonances in the $\Lambda_c^+ K^-$ channel, $\Xi_c(2923)^0$, $\Xi_c(2939)^0$ and $\Xi_c(2965)^0$ \cite{PhysRevLett.124.222001} which were interpreted as $1P$-wave states of the flavor sextet \cite{PhysRevD.105.074029,EOP2023}. The Belle Collaboration studied the electromagnetic decay of the excited charm baryons $\Xi_c(2790)$ and $\Xi_c(2815)$ \cite{PhysRevD.102.071103} which led to the assignment of these resonances as $1P$-wave states of the flavor anti-triplet \cite{Wang2017,PhysRevD.105.074029,EOP2023}. 
The Belle Collaboration determined the spin and parity of the charmed-strange baryon $\Xi_c(2970)^+$ to be $J^P=1/2^+$ \cite{MoonPRD103} which in the quark model would correspond to two quanta of excitation.  

As an additional remark, we note that $\Xi_c(2923)$, $\Xi_c(2930)$ and $\Xi_c(2936)$ were omitted from the Particle Data Group Summary Table \cite{PDG26}. 

As for the baryons containing one bottom quark, the early observations of LHCb lead to $\Xi^-_b(5935)$ and $\Xi^-_b(5955)$ resonances close to the threshold  $\Xi_b\pi$ \cite{PhysRevLett.114.062004}, which in the quark model are typically assigned with spin-parity $J^P=1/2^+$ and $3/2^+$ prime ground states, respectively. For unprimed resonances $\Xi_b$ and $\Xi^*_b$ \cite{PhysRevLett.113.242002,PhysRevD.99.052006} we assign the same quantum numbers. However, the primary difference from our previous assignment resides in the multiplets to which these baryons belong.

Recently, the LCHb Collaboration confirmed the existence of the $\Xi_b(6100)^-$ baryon and announced the discovery of $\Xi_b(6087)^0$ and $\Xi_b(6095)^0$ in the $\Xi_b^0 \pi^+ \pi^-$ system \cite{lhcbcollaboration2023observation}. The $\Xi_b(6095)^0$ and $\Xi_b(6100)^-$ resonances were  interpreted as members of the isospin doublet $^2\lambda(\Xi_b)_{J=3/2}$, whereas the $\Xi_b(6087)^0$ baryon was assigned as $^2\lambda(\Xi_b)_{J=1/2}$ \cite{EOP2023}. 
In addition, the LHCb Collaboration announced the discovery of $\Xi_b(6327)^0$ and $\Xi_b(6333)^0$ which were assigned as $1D$-wave states \cite{PhysRevLett.128.162001}. 

In the present analysis we limit ourselves to ground state $1S$-wave and excited $1P$-wave states. Table~\ref{XiQ} shows the partial and total decay widths of $S$- and $P$-wave $\Xi_Q$ and $\Xi'_Q$ baryons in the EEM. All calculated widths are relatively small, the largest width being 28.4 MeV for the $^4\rho(\Xi_c)_{5/2}$ anti-triplet configuration. A comparison with the available experimental data shows a reasonable agreement both in the charm and in the bottom sector. 

\begin{table*}[t]
\caption{Partial and total decay widths of sextet $\Xi'_Q$ and anti-triplet $\Xi_Q$ baryons in MeV.}
\label{XiQ}
\centering
\begingroup
\footnotesize
\setlength{\tabcolsep}{1.2pt}
\renewcommand{\arraystretch}{0.94}
\begin{adjustbox}{max width=\textwidth,center}
\begin{tabular}{ccrrrrrrrrrcl}
\noalign{\smallskip}
\toprule
& \multicolumn{8}{c}{Present work}  &&&& \\ 
\noalign{\smallskip}
\cmidrule(lr){2-9} 
\noalign{\smallskip}
State & $M^{th}(\text{MeV})$& $\Sigma_c\overline{K}$ & $\Xi'_c\pi$ & $\Sigma^*_c\overline{K}$ & $\Xi'^*_c\pi$ & $\Lambda_c\overline{K}$ & $\Xi_c\pi$ & $\Gamma_{\text{EEM}}$ 
& $^3P_0$ \cite{PhysRevD.107.034031} & $\chi$QM \cite{Wang2017} & $\Gamma_{\text{exp}}$ \cite{PDG26,PhysRevLett.124.222001} & Baryon \\
\noalign{\smallskip}
\hline
\noalign{\smallskip}
$^2(\Xi'_{c})_{1/2}$       & $2586 \pm 2$  &  -- &  -- &  -- &  -- &  -- &   -- &   -- & & & & $\Xi'^+_{c}$, $\Xi'^0_{c}$ \\
$^4(\Xi'_{c})_{3/2}$       & $2653 \pm 2$  &  -- &  -- &  -- &  -- &  -- &  2.6 &  2.6 &   0.4 &      & $2.14 \pm 0.19$ & $\Xi_{c}(2645)^+$ \\
                          &    &  -- &  -- &  -- &  -- &  -- &  2.6 &  2.6 &   0.4 &      & $2.35 \pm 0.22$ & $\Xi_{c}(2645)^0$ \\
$^2\lambda(\Xi'_{c})_{1/2}$ & $2890 \pm 2$ &  -- & 0.6 &  -- & 0.1 & 0.4 &  2.4 &  3.6 &   7.3 & 21.67 & & \\
$^2\lambda(\Xi'_{c})_{3/2}$ & $2922 \pm 2$ &  -- & 1.8 &  -- & 0.2 & 1.8 &  4.7 &  8.5 &  27.9 & 20.89 & $14.8 \pm 9.1$ & $\Xi_{c}(2930)^+$ \\
                           &   &  -- & 1.8 &  -- & 0.2 & 1.8 &  4.7 &  8.5 &  27.9 & 20.89 & $10.2 \pm 1.4$ & $\Xi_{c}(2930)^0$ \\
$^4\lambda(\Xi'_{c})_{1/2}$ & $2923 \pm 2$ &  -- & 0.6 &  -- & 0.1 & 2.0 &  6.9 &  9.6 &   5.1 & 37.05 &  $5.3 \pm 1.7$ & $\Xi_{c}(2923)^+$ \\
& &  -- & 0.6 &  -- & 0.1 & 2.0 &  6.9 &  9.6 &   5.1 & 37.05 &  $5.8 \pm 1.3$ & $\Xi_{c}(2923)^0$ \\
$^4\lambda(\Xi'_{c})_{3/2}$& $2956 \pm 2$  & 0.0 & 0.1 &  -- & 1.6 & 0.6 &  1.3 &  3.6 &  18.9 & 12.33 & $14.1 \pm 1.6$ & $\Xi_{c}(2965)^0$ \\
$^4\lambda(\Xi'_{c})_{5/2}$ & $3011 \pm 2$ & 0.3 & 1.6 &  -- & 2.5 & 6.3 & 11.2 & 21.9 &  43.1 & 20.20 &                & \\
$^2\rho(\Xi'_{c})_{1/2}$    &  $3029 \pm 2$  & 0.5 & 3.5 & 0.1 & 4.7 &   0 &    0 &  8.8 & 156.6 &       &  $7.8 \pm 1.9$ & $\Xi_{c}(3055)^+$ \\
$^2\rho(\Xi'_{c})_{3/2}$  & $3062 \pm 2$   & 3.5 & 6.4 & 0.5 & 5.9 &   0 &    0 & 16.2 &  99.9 &       &  $3.6 \pm 1.1$ & $\Xi_{c}(3080)^+$ \\
                           &   & 3.5 & 6.4 & 0.5 & 5.9 &   0 &    0 & 16.2 &  99.9 &       &  $5.6 \pm 2.2$ & $\Xi_{c}(3080)^0$ \\
\noalign{\smallskip}
\hline
\noalign{\smallskip}
$^2(\Xi_{c})_{1/2}$      & $2474 \pm 2$   &  -- &  -- &  -- &  -- &   -- &   -- &   -- & & & & $\Xi_{c}^+$, $\Xi_{c}^0$ \\
$^2\lambda(\Xi_{c})_{1/2}$& $2917 \pm 2$  &  -- & 0.0 &  -- &  -- &   -- &    0 &  0.0 &   2.6 & 3.61 &  $8.9 \pm 1.0$  & $\Xi_{c}(2790)^+$ \\
                           &  &  -- & 0.0 &  -- &  -- &   -- &    0 &  0.0 &   2.6 & 3.61 & $10.0 \pm 1.1$  & $\Xi_{c}(2790)^0$ \\
$^2\lambda(\Xi_{c})_{3/2}$ & $2810 \pm 2$ &  -- & 0.1 &  -- & 0.0 &    0 &    0 &  0.1 &   4.5 & 2.11 & $2.43 \pm 0.26$ & $\Xi_{c}(2815)^+$ \\
                           &  &  -- & 0.1 &  -- & 0.0 &    0 &    0 &  0.1 &   4.5 & 2.11 & $2.54 \pm 0.25$ & $\Xi_{c}(2815)^0$ \\
$^2\rho(\Xi_{c})_{1/2}$   & $2950 \pm 2$  &  -- & 1.0 &  -- & 0.2 &  0.9 &  3.3 &  5.5 &  17.0 & & & \\
$^2\rho(\Xi_{c})_{3/2}$   & $2950 \pm 2$  & 0.0 & 2.7 &  -- & 0.4 &  2.8 &  6.1 & 11.9 &  89.1 & & & \\
$^4\rho(\Xi_{c})_{1/2}$   & $2983 \pm 2$  & 0.0 & 0.9 &  -- & 0.2 &  3.4 &  9.2 & 13.7 &  12.9 & & & \\
$^4\rho(\Xi_{c})_{3/2}$   & $3038 \pm 2$  & 0.0 & 0.2 &  -- & 2.7 &  0.8 &  1.6 &  5.3 &  56.2 & & & \\
$^4\rho(\Xi_{c})_{5/2}$   &  $2777 \pm 2$  & 0.8 & 2.1 & 0.1 & 3.7 &  8.2 & 13.6 & 28.4 & 122.1 & & & \\
\noalign{\smallskip}
\hline
\noalign{\smallskip}
& \multicolumn{8}{c}{Present work}  &&&& \\ 
\noalign{\smallskip}
\cmidrule(lr){2-9} 
\noalign{\smallskip}
State &$M^{th}(\text{MeV})$& $\Sigma_b\overline{K}$ & $\Xi'_b\pi$ & $\Sigma^*_b\overline{K}$ & $\Xi'^*_b\pi$ & $\Lambda_b\overline{K}$ & $\Xi_b\pi$ & $\Gamma_{\text{EEM}}$ & $^3P_0$ \cite{garciatecocoatzi2023decay} & $\chi$QM \cite{Wang2017} & $\Gamma_{\text{exp}}$ \cite{PDG26,lhcbcollaboration2023observation} & Baryon \\
\noalign{\smallskip}
\hline
\noalign{\smallskip}
$^2(\Xi'_{b})_{1/2}$      & $5935 \pm 2$   &  -- &  -- &  -- &  -- &  -- &  0.0 &  0.0 & & 0.08  & $0.03 \pm 0.03$ & $\Xi'_{b}(5935)^-$ \\
$^4(\Xi'_{b})_{3/2}$      & $5957 \pm 2$    &  -- &  -- &  -- &  -- &  -- &  1.1 &  1.1 &   0.2 & 0.73  & $0.87 \pm 0.08$ & $\Xi_{b}(5945)^0$ \\
&  &  -- &  -- &  -- &  -- &  -- &  1.1 &  1.1 &   0.2 & 1.23  & $1.43 \pm 0.11$ & $\Xi_{b}(5955)^-$ \\
$^2\lambda(\Xi'_{b})_{1/2}$ & $6201 \pm 3$ &  -- & 0.4 &  -- & 0.2 & 0.5 &  3.7 &  4.8 &   3.0 & 27.05 & & \\
$^2\lambda(\Xi'_{b})_{3/2}$& $6207 \pm 3$    &  -- & 0.9 &  -- & 0.2 & 1.4 &  5.6 &  8.0 &  29.5 & 24.15 & & \\
$^4\lambda(\Xi'_{b})_{1/2}$& $6216 \pm 3$  &  -- & 0.3 &  -- & 0.2 & 1.6 &  8.8 & 10.9 &   3.7 & 32.24 & & \\
$^4\lambda(\Xi'_{b})_{3/2}$& $6223 \pm 3$  &  -- & 0.1 &  -- & 1.1 & 0.4 &  1.3 &  2.9 &   7.6 & 15.83 & & \\
$^4\lambda(\Xi'_{b})_{5/2}$ & $6234 \pm 3$ &  -- & 0.4 &  -- & 1.1 & 3.0 &  8.7 & 13.2 &  31.4 & 24.39 & $18.6^{+1.5}_{-1.6}$ & $\Xi_{b}(6227)^0$ \\
&  &  -- & 0.4 &  -- & 1.1 & 3.0 &  8.7 & 13.2 &  31.4 & 24.39 & $19.9 \pm 2.6$ & $\Xi_{b}(6227)^-$ \\
$^2\rho(\Xi'_{b})_{1/2}$  & $6366 \pm 3$   & 0.2 & 3.5 & 0.7 & 7.8 &   0 &    0 & 12.3 & 196.6 & & & \\
$^2\rho(\Xi'_{b})_{3/2}$ &  $6373 \pm 3$   & 1.3 & 5.1 & 0.6 & 7.2 &   0 &    0 & 14.2 &  97.2 & & & \\
\noalign{\smallskip}
\hline
\noalign{\smallskip}
$^2(\Xi_{b})_{1/2}$   &  $5812 \pm 2$     &  -- &  -- &  -- &  -- &  -- &   -- &   -- & &      &                 & $\Xi_{b}^0$, $\Xi_{b}^-$ \\
$^2\lambda(\Xi_{b})_{1/2}$& $6077 \pm 3$  &  -- & 0.0 &  -- &  -- &  -- &    0 &  0.0 &  0.2 & 2.88 & $2.43 \pm 0.52$ & $\Xi_{b}(6087)^0$ \\
$^2\lambda(\Xi_{b})_{3/2}$& $6084 \pm 3$  &  -- & 0.0 &  -- &  -- &  -- &    0 &  0.0 &  1.1 & 2.95 & $0.94 \pm 0.31$ & $\Xi_{b}(6100)^-$ \\
&  &  -- & 0.0 &  -- &  -- &  -- &    0 &  0.0 &  1.1 & 2.95 & $0.50 \pm 0.35$ & $\Xi_{b}(6095)^0$ \\
$^2\rho(\Xi_{b})_{1/2}$  & $6243 \pm 3$   &  -- & 0.7 &  -- & 0.4 & 1.2 &  4.2 &  6.5 &  8.7 & & & \\
$^2\rho(\Xi_{b})_{3/2}$& $6249 \pm 3$     &  -- & 1.4 &  -- & 0.4 & 2.4 &  5.9 & 10.1 & 65.9 & & & \\
$^4\rho(\Xi_{b})_{1/2}$ & $6258 \pm 3$    &  -- & 0.5 &  -- & 0.3 & 3.3 &  9.6 & 13.7 &  6.0 & & & \\
$^4\rho(\Xi_{b})_{3/2}$ & $6265 \pm 3$    &  -- & 0.1 &  -- & 2.0 & 0.6 &  1.4 &  4.1 & 26.1 & & & \\
$^4\rho(\Xi_{b})_{5/2}$ & $6276 \pm 3$    &  -- & 0.6 &  -- & 1.7 & 4.2 &  8.8 & 15.4 & 68.5 & & & \\
\noalign{\smallskip}
\botrule
\end{tabular}
\end{adjustbox}
\endgroup
\end{table*}

\subsection{$\Omega_Q$ baryons}

\begin{table*}[t]
\centering
\caption{Partial and total decay widths of sextet $\Omega_Q$ baryons in MeV.}
\label{OmegaQ}
\begingroup
\setlength{\tabcolsep}{1.4pt}
\renewcommand{\arraystretch}{0.96}
\begin{adjustbox}{max width=\textwidth,center}
\begin{tabular}{ccrrrrrrcl}
\noalign{\smallskip}
\toprule
& \multicolumn{5}{c}{Present work}  &&&& \\ 
\noalign{\smallskip}
\cmidrule(lr){2-6} 
\noalign{\smallskip}
State & $M^{th}(\text{MeV})$& $\Xi'_c \overline{K}$ & $\Xi'^*_c \overline{K}$ & $\Xi_c \overline{K}$ & $\Gamma_{\text{EEM}}$ & $^3P_0$ \cite{PhysRevD.107.034031} & $\chi$QM \cite{WangOme} & $\Gamma_{\text{exp}}$ \cite{PDG26,PhysRevLett.131.131902} & Baryon \\
\noalign{\smallskip}
\hline
\noalign{\smallskip}
$^2(\Omega_{c})_{1/2}$       & $2733 \pm 2$  &  -- &  -- &   -- &   -- &  
& & & $\Omega_{c}^0$ \\
$^4(\Omega_{c})_{3/2}$       & $2799 \pm 2$  &  -- &  -- &   -- &   -- &  
& & & $\Omega_{c}(2770)^0$ \\
$^2\lambda(\Omega_{c})_{1/2}$& $3016 \pm 2$   &  -- &  -- &  0.0 &  0.0 &  4.1 
& 4.38/4.28 & $3.83^{+1.61}_{-0.37}$ & $\Omega_{c}(3000)^0$ \\
$^2\lambda(\Omega_{c})_{3/2}$ & $3048 \pm 2$ &  -- &  -- &  1.8 &  1.8 & 26.3 
& 4.96      & $3.4^{+0.7}_{-0.8}$ & $\Omega_{c}(3065)^0$ \\
$^4\lambda(\Omega_{c})_{1/2}$& $3049 \pm 2$  &  -- &  -- &  0.9 &  0.9 &  7.6 
&  --       & $<1.8$                 & $\Omega_{c}(3050)^0$\\
$^4\lambda(\Omega_{c})_{3/2}$&  $3082 \pm 2$  & 0.0 &  -- &  0.8 &  0.8 &  6.7 
& 0.94      & $8.48^{+0.75}_{-1.68}$ & $\Omega_{c}(3090)^0$ \\
$^4\lambda(\Omega_{c})_{5/2}$& $3136 \pm 2$  & 0.4 &  -- & 12.3 & 12.7 & 50.1 
& 9.53      & $<2.5$                 & $\Omega_{c}(3120)^0$ \\
$^2\rho(\Omega_{c})_{1/2}$ & $3131 \pm 2$     & 0.1 &  -- &   0  &  0.1 & 14.4 & & & \\
$^2\rho(\Omega_{c})_{3/2}$ & $3164 \pm 2$    & 2.5 & 0.1 &   0  &  2.7 & 71.6 
&           & $50^{+12}_{-21}$ & $\Omega_{c}(3185)^0$ \\
\noalign{\smallskip}
\hline
\noalign{\smallskip}
& \multicolumn{5}{c}{Present work}  &&&& \\ 
\noalign{\smallskip}
\cmidrule(lr){2-6} 
\noalign{\smallskip}
State&$M^{th}(\text{MeV})$ & $\Xi'_b \overline{K}$ & $\Xi'^*_b \overline{K}$ & $\Xi_b \overline{K}$ & $\Gamma_{\text{EEM}}$
& $^3P_0$ \cite{garciatecocoatzi2023decay} & $\chi$QM \cite{Wang2017} & $\Gamma_{\text{exp}}$ \cite{PDG26,PhysRevLett.124.082002} & Baryon \\
\noalign{\smallskip}
\hline
\noalign{\smallskip}
$^2(\Omega_{b})_{1/2}$      & $6078 \pm 2$   &  -- &  -- &  -- &  -- & 
& & & $\Omega_{b}^-$ \\
$^4(\Omega_{b})_{3/2}$      & $6100 \pm 3$   &  -- &  -- &  -- &  -- & & & & \\
$^2\lambda(\Omega_{b})_{1/2}$& $6321 \pm 2$   &  -- &  -- & 0.0 & 0.0 &  4.6 
& 49.38 & $<4.2$ & $\Omega_b(6316)^-$ \\
$^2\lambda(\Omega_{b})_{3/2}$& $6328 \pm 2$  &  -- &  -- & 0.6 & 0.6 & 24.0 
&  1.82 & $<4.7$ & $\Omega_b(6330)^-$ \\
$^4\lambda(\Omega_{b})_{1/2}$ & $6337 \pm 3$ &  -- &  -- & 0.1 & 0.1 & 10.7 
& 94.98 & $<1.8$ & $\Omega_b(6340)^-$ \\
$^4\lambda(\Omega_{b})_{3/2}$& $6343 \pm 3$  &  -- &  -- & 0.3 & 0.3 &  6.3 
&  0.22 & $< 3.2$ & $\Omega_b(6350)^-$ \\
$^4\lambda(\Omega_{b})_{5/2}$& $6354 \pm 3$  &  -- &  -- & 2.7 & 2.7 & 40.5 &  1.60 & & \\
$^2\rho(\Omega_{b})_{1/2}$   & $6467 \pm 3$  & 0.0 & 0.1 &  0  & 0.1 &  9.8 & & & \\
$^2\rho(\Omega_{b})_{3/2}$  & $6474 \pm 3$  & 0.6 & 0.2 &  0  & 0.8 & 53.5 & & & \\
\noalign{\smallskip}
\botrule
\end{tabular}
\end{adjustbox}
\endgroup
\end{table*}

In 2017, the LHCb Collaboration  announced the observation of five resonances $\Omega_c(3000)^0$, $\Omega_c(3050)^0$, $\Omega_c(3065)^0$, $\Omega_c(3090)^0$ and $\Omega_c(3119)^0$ \cite{PhysRevLett.118.182001}. At the same time, a further resonance was reported $\Omega_c(3188)^0$, but because of the absence of sufficient statistic, it was not claimed as an authentic resonance. All these signals were discovered in the $\Xi_c^+K^-$ decay channel. One year later, the Belle Collaboration confirmed the observation of the first four resonances together with $\Omega_c(3188)^0$, however, they did not observe the $\Omega_c(3119)^0$ \cite{PhysRevD.97.051102}. 
In a more recent study by the LHCb Collaboration the existence of the five resonances was confirmed, together with the observation of two new states, $\Omega_c(3185)^0$ and $\Omega_c(3327)^0$ \cite{PhysRevLett.131.131902}. 
 
The experimental knowledge on $\Omega_b$ has been expanded by the LHCb collaboration who reported 4 signals in the $\Xi_b^0K^-$ mass spectrum corresponding to $\Omega_b(6316)^-$, $\Omega_b(6330)^-$, $\Omega_b(6340)^-$ and $\Omega_b(6350)^-$ \cite{PhysRevLett.124.082002}. 

The ground state $S$-wave $\Omega_Q$ baryons have been observed with the exception of the $^4(\Omega_b)_{3/2}$ state. In the charm sector, six of the seven $P$-wave states have been observed, as well as four of the seven in the bottom sector. The partial widths of the $^2\rho(\Omega_Q)_J$ configurations to the $\Xi_Q \overline{K}$ channel are forbidden. The calculated widths in the EEM are small, $\lesssim 12$ MeV, in agreement with the experimental data \cite{luo2026quantumnumbersexcitedxicprime}. 

\subsection{$\Xi_{QQ}$ and $\Omega_{QQ}$ baryons}

In this section we present the results for the doubly-heavy baryons, $\Xi_{QQ}$ and $\Omega_{QQ}$. 
The LHCb Collaboration has identified candidates for all three members of the flavor triplet, $\Xi^{++}_{cc}(3621)$ \cite{PhysRevLett.119.112001}, $\Xi^{+}_{cc}$ ($dcc$) \cite{dmv6-7gdv} and $\Omega_{cc}^{+}$ ($scc$)~\cite{Wang:2026OmegaCC1,LHCb:2026OmegaCC2} which can be assigned as the ground state $S$-wave baryons with $J^P=1/2^+$ \cite{EOP2023}. Currently, there is a considerable research effort in the hadron physics community to analyze and characterize the properties of doubly-heavy baryons \cite{PhysRevD.81.094505,PhysRevD.86.094504,PhysRevD.87.094512,PhysRevD.90.074504,PhysRevD.90.094507,PhysRevD.92.034504,PhysRevD.96.034511,PhysRevD.90.074501}. 

The strong decay widths can be obtained in the same way as for singly-heavy baryons by interchanging the role of the light and heavy quarks, {\it i.e.} interchanging $m$ and $m'$. As a consequence, for doubly-heavy baryons the frequency of the $\rho$ mode is less than that of the $\lambda$ mode. 

Tables~\ref{XiQQ} and \ref{OmegaQQ} show that in contrast to the chiral quark model \cite{PhysRevD.96.094005} and the constituent quark model \cite{arXiv:2408.11578} the EEM decay widths are very small, $\lesssim 1$ MeV. The strong decay of the $^{2}\rho$ configuration into another doubly heavy baryon and an octet meson is forbidden by the structure of the spin-flavor wave functions, whereas the strong decays of the $^{2}\lambda$ and $^{4}\lambda$ configurations are suppressed in the present approach by the mass dependent factor in Eqs.~(\ref{ulambda}) and (\ref{tlambda}).

\begin{table*}[t]
\centering
\caption{Partial and total decay widths of triplet $\Xi_{QQ}$ baryons in MeV.}
\label{XiQQ}
\begingroup
\setlength{\tabcolsep}{2.8pt}
\renewcommand{\arraystretch}{1.00}
\begin{adjustbox}{max width=\textwidth,center}
\begin{tabular}{ccrrrrr}
\noalign{\smallskip}
\toprule
& \multicolumn{4}{c}{Present work} & \cite{PhysRevD.96.094005} & \cite{arXiv:2408.11578} \\ 
\cmidrule(lr){2-5} 
\noalign{\smallskip}
State &$M^{th}(\text{MeV})$& $\Xi_{cc}\pi$ & $\Xi^*_{cc}\pi$ & $\Gamma_{\mathrm{EEM}}$ & $\chi$QM & CQM \\
\noalign{\smallskip}
\hline 
\noalign{\smallskip}
$^2(\Xi_{cc})_{1/2}$       & $3619 \pm 2$  &  -- &  -- &  -- & & \\
$^4(\Xi_{cc})_{3/2}$       & $3686 \pm 2$  &  -- &  -- &  -- & & \\
$^2\rho(\Xi_{cc})_{1/2}$   & $3823 \pm 2$  &   0 &  -- & 0.0 & & \\
$^2\rho(\Xi_{cc})_{3/2}$   & $3855 \pm 2$  &   0 &   0 & 0.0 & & \\
$^2\lambda(\Xi_{cc})_{1/2}$& $4048 \pm 2$  & 0.0 & 0.1 & 0.1 & 49.5 & 440.4 \\
$^2\lambda(\Xi_{cc})_{3/2}$& $4081 \pm 2$  & 0.1 & 0.1 & 0.2 & 122.6 & 336.6 \\
$^4\lambda(\Xi_{cc})_{1/2}$& $4082 \pm 2$  & 0.1 & 0.0 & 0.1 & 134.2 &  53.9 \\
$^4\lambda(\Xi_{cc})_{3/2}$ & $4114 \pm 2$  & 0.1 & 0.2 & 0.2 & 92.2 &  34.7 \\
$^4\lambda(\Xi_{cc})_{5/2}$ & $4169 \pm 2$  & 0.7 & 0.3 & 1.0 & 98.1 &  92.2 \\
\noalign{\smallskip} 
\hline
\noalign{\smallskip}
& \multicolumn{4}{c}{Present work} & \cite{PhysRevD.96.094005} & \cite{arXiv:2408.11578} \\ 
\cmidrule(lr){2-5} 
\noalign{\smallskip}
State &$M^{th}(\text{MeV})$& $\Xi_{bb}\pi$ & $\Xi^*_{bb}\pi$ & $\Gamma_{\mathrm{EEM}}$ & $\chi$QM & CQM \\
\noalign{\smallskip}
\hline 
\noalign{\smallskip}
$^2(\Xi_{bb})_{1/2}$       & $10295 \pm 3$  &  -- &  -- &  -- & & \\
$^4(\Xi_{bb})_{3/2}$       & $10317 \pm 4$  &  -- &  -- &  -- & & \\
$^2\rho(\Xi_{bb})_{1/2}$   & $10411 \pm 3$  &  -- &  -- &  -- & & \\
$^2\rho(\Xi_{bb})_{3/2}$   & $10417 \pm 3$  &   0 &  -- & 0.0 & & \\
$^2\lambda(\Xi_{bb})_{1/2}$& $10700 \pm 4$  & 0.0 & 0.0 & 0.0 & 81.4 & 422.4 \\
$^2\lambda(\Xi_{bb})_{3/2}$& $10707 \pm 4$  & 0.0 & 0.0 & 0.1 & 71.7 & 384.1 \\
$^4\lambda(\Xi_{bb})_{1/2}$& $10716 \pm 4$  & 0.0 & 0.0 & 0.0 & 26.7 &  45.1 \\
$^4\lambda(\Xi_{bb})_{3/2}$& $10722 \pm 4$  & 0.0 & 0.1 & 0.1 & 47.6 &  39.4 \\
$^4\lambda(\Xi_{bb})_{5/2}$& $10733 \pm 4$  & 0.1 & 0.1 & 0.2 & 95.1 & 108.3 \\
\noalign{\smallskip}
\botrule
\end{tabular}
\end{adjustbox}
\endgroup 
\end{table*}

\begin{table*}[t]
\centering
\caption{Partial and total decay widths of triplet $\Omega_{cc}$ baryons in MeV.}
\label{OmegaQQ}
\begingroup
\setlength{\tabcolsep}{2.8pt}
\renewcommand{\arraystretch}{1.00}
\begin{adjustbox}{max width=\textwidth,center}
\begin{tabular}{ccrrrrr}
\noalign{\smallskip}
\toprule
& \multicolumn{4}{c}{Present work} & \cite{PhysRevD.96.094005} & \cite{arXiv:2408.11578} \\ 
\cmidrule(lr){2-5} 
\noalign{\smallskip}
State&$M^{th}(\text{MeV})$ & $\Xi_{cc}\overline{K}$ & $\Xi^*_{cc}\overline{K}$ & $\Gamma_{\mathrm{EEM}}$ & $\chi$QM & CQM \\
\noalign{\smallskip}
\hline 
\noalign{\smallskip}
$^2(\Omega_{cc})_{1/2}$       & $3766 \pm 2$  &  -- &  -- &  -- & & \\
$^4(\Omega_{cc})_{3/2}$       & $3833 \pm 2$  &  -- &  -- &  -- & & \\
$^2\rho(\Omega_{cc})_{1/2}$   & $3970 \pm 2$  &  -- &  -- &  -- & & \\
$^2\rho(\Omega_{cc})_{3/2}$   & $4002 \pm 2$  &  -- &  -- &  -- & & \\
$^2\lambda(\Omega_{cc})_{1/2}$& $4111 \pm 2$  &  -- &  -- &  -- & 35.5 & 450.9 \\
$^2\lambda(\Omega_{cc})_{3/2}$& $4144 \pm 2$  & 0.0 &  -- & 0.0 & 185.4 &   0.0 \\
$^4\lambda(\Omega_{cc})_{1/2}$& $4145 \pm 2$  & 0.1 &  -- & 0.1 & 323.0 &   0.0 \\
$^4\lambda(\Omega_{cc})_{3/2}$& $4177 \pm 2$  & 0.0 &  -- & 0.0 & 140.1 &   0.5 \\
$^4\lambda(\Omega_{cc})_{5/2}$& $4232 \pm 2$  & 0.2 & 0.0 & 0.3 & 45.9 &   1.5 \\
\noalign{\smallskip}
\hline
\noalign{\smallskip}
& \multicolumn{4}{c}{Present work} & \cite{PhysRevD.96.094005} & \cite{arXiv:2408.11578} \\ 
\cmidrule(lr){2-5} 
\noalign{\smallskip}
State&$M^{th}(\text{MeV})$ & $\Xi_{bb}\overline{K}$ & $\Xi^*_{bb}\overline{K}$ & $\Gamma_{\mathrm{EEM}}$ & $\chi$QM & CQM \\
\noalign{\smallskip}
\hline 
\noalign{\smallskip}
$^2(\Omega_{bb})_{1/2}$       & $10438 \pm 3$  &  -- &  -- &  -- & & \\
$^4(\Omega_{bb})_{3/2}$       & $10460 \pm 4$  &  -- &  -- &  -- & & \\
$^2\rho(\Omega_{bb})_{1/2}$   & $10554 \pm 3$  &  -- &  -- &  -- & & \\
$^2\rho(\Omega_{bb})_{3/2}$   & $10560 \pm 3$  &  -- &  -- &  -- & & \\
$^2\lambda(\Omega_{bb})_{1/2}$& $10762 \pm 4$  &  -- &  -- &  -- & 30.3 & 213.4 \\
$^2\lambda(\Omega_{bb})_{3/2}$& $10768 \pm 4$  & 0.0 &  -- & 0.0 & 89.1 &   0.0 \\
$^4\lambda(\Omega_{bb})_{1/2}$& $10778 \pm 4$  & 0.2 &  -- & 0.2 &149.4 &   0.0 \\
$^4\lambda(\Omega_{bb})_{3/2}$& $10784 \pm 4$  & 0.0 & 0.2 & 0.2 & 99.2 &   0.1 \\
$^4\lambda(\Omega_{bb})_{5/2}$& $10795 \pm 4$  & 0.0 & 0.0 & 0.1 & 25.0 &   0.5 \\
\noalign{\smallskip}
\botrule
\end{tabular}
\end{adjustbox}
\endgroup
\end{table*}

\section{Discussion}
\label{discussion}

In Tables~\ref{SigmaQ}-\ref{OmegaQQ} we made a comparison of the EEM results with other model calculations, in particular with the chiral quark model ($\chi$QM) \cite{Wang2017,WangOme} and the $^3P_0$ model \cite{PhysRevD.107.034031,garciatecocoatzi2023decay}, both of which present a comprehensive study of (almost all) $S$- and $P$-wave singly-heavy baryons of the sextet, $\Sigma_Q$, $\Xi'_Q$ and $\Omega_Q$, and the anti-triplet, $\Lambda_Q$ and $\Xi_Q$, and doubly-heavy baryons of the triplet, $\Xi_{QQ}$ and $\Omega_{QQ}$. 

All three approaches share the use of the wave functions of the harmonic oscillator quark model. In all three cases the relative flavor couplings are given by the flavor $SU(3)$ isoscalar factors presented in section~\ref{widths}. Note that there are many differences with the relative flavor couplings of Ref.~\cite{PhysRevD.107.034031} (apart from the phase conventions used for the isoscalar factors, Baird-Biedenharn and De Swart, respectively). 

The main difference between the three approaches is in the decay model used to calculate the decay widths. In the present study we have adopted a harmonic oscillator quark model and used the corresponding wave functions to calculate the decay widths in the elementary emission model in which we used the same values of the coefficients, $g$ and $h$ for all decays. In Refs.~\cite{PhysRevD.107.034031,garciatecocoatzi2023decay} the harmonic oscillator quark model is combined with a $^3P_0$ model for the two-body decays of heavy baryons. 

In the $\chi$QM of \cite{Wang2017,WangOme} a hybrid approach is used in which the masses are taken from a relativistic quark-diquark calculation \cite{PhysRevD.84.014025} in combination with harmonic oscillator wave functions. The strong coupling is derived by taking the nonrelativistic limit of the effective quark-pseudoscalar-meson interaction in the chiral quark model. The final form of the strong coupling is very similar to the one used in the elementary emission model, see {\it e.g.} \cite{LeYaouanc:1988fx,PhysRevD.21.1868,BIJKER200089}, but with the difference that the coefficients $g$ and $h$ depend on the energy of the final hadrons. The relative sign of $g$ and $h$ is the same. Finally, the decay widths are calculated by adding an extra factor that depends on the strength of the quark-meson couplings and the pion and kaon decay constants. 

A comparison of the predictions of the three models for the singly-heavy baryons in Tables~\ref{SigmaQ}-\ref{OmegaQ} shows that there is a qualitative agreement for the experimentally known decays with the exception of the decay widths of $\Xi_c(3055)$ and $\Xi_c(3080)$ which are overpredicted by a large amount in the $^3P_0$ model. For the unknown decays the results differ wildly, at times by an order of magnitude. In general, the $^3P_0$ model predicts much larger widths than the EEM and $\chi$QM. 

The predictions of the decay widths for the doubly-heavy baryons in Tables~\ref{XiQQ} and \ref{OmegaQQ} show a very small width for the EEM and a much larger value of the order of 100s MeV for the $\chi$QM \cite{PhysRevD.96.094005} and the constituent quark model \cite{arXiv:2408.11578}. 

In general, the experimental information is somewhat more limited in the bottom sector compared to that in the charm sector as can be easily seen across the Tables~\ref{SigmaQ}-\ref{OmegaQQ}. This is expected given  the higher production rates of charm baryons  in high-energy collision experiments than bottom baryons, as they are produced more frequently due to lower mass thresholds \cite{PDG26}. Additionally, charm baryons often have more accessible decay channels, making them easier to study, while bottom baryons typically decay through more complex processes that yield fewer observable final states.

\section{Summary and conclusions}
\label{summary}

In this contribution, we presented the results of a simultaneous global analysis of two-body strong decay widths of singly and doubly heavy baryons, $\Sigma_Q$, $\Xi'_Q$, $\Omega_Q$, $\Lambda_Q$, $\Xi_Q$ y $\Xi_{QQ}$, $\Omega_{QQ}$, respectively, with $Q=c,b$. The baryon masses and corresponding wave functions were obtained from a previous study in the framework of a non-relativistic harmonic oscillator quark model \cite{EOP2023}. For the strong couplings we adopted the elementary emission model for the two-body decays into a heavy baryon and a light pseudo-scalar meson. The calculations were limited to the ground state $1S$-wave and excited state $1P$-wave baryons. 

A comparison with the available experimental information shows a reasonable agreement with the data. In the vast majority of cases, the estimated strong decay widths in the elementary emission model are smaller than the observed experimental widths across different charm and bottom baryon states. The theoretical values correspond exclusively to two-body decay channels, whereas the experimental widths incorporate other decay channels as well. 

In addition, a comparison of the decay widths of singly-heavy baryons with calculations in the $^{3}P_0$ model and the chiral quark model shows, with the exception of only a few cases, a qualitative agreement for the experimentally known decays. However, for the unknown decays the results differ wildly, at times by an order of magnitude. In general, the $^3P_0$ model predicts much larger widths than the EEM and $\chi$QM. 

The decay widths for the doubly-heavy baryons in the elementary emission model are either forbidden or suppressed by quark mass dependent factors in the radial integrals. In contrast, the chiral quark model and the constituent quark model give much larger values, in many cases of the order of 100s MeV. 

In conclusion, in recent years a wealth of new information on heavy baryons has been obtained by several experimental collaborations, especially for masses and decay widths. With the exception of $\Xi_c(2970)$ and $\Xi_c(3055)$ for which the spin and parity have been measured, the assignment of quantum numbers of mostly based on systematics guided by quark model considerations. For the interpretation of the nature of heavy baryon systems and the assignment of quantum numbers more experimental information is needed on radiative and strong decay widths, as well as spin and parity measurements. In this manuscript and in a previous publication \cite{EOP2023} we presented a global analysis of heavy baryons including mass spectra and electromagnetic and strong decay widths which may help to interpret heavy baryon resonances and identify unknown states which based on the quark model are expected to exist but so far have not been found experimentally. 

\bmhead{Acknowledgements}
This work was supported in part by PAPIIT-DGAPA (UNAM, Mexico) grant No. IN115426. E.O.-P. acknowledges support from the Momentum
(Lend\"ulet) Programme of the Hungarian Academy of Sciences
under Grant No.~LP2025-13/2025.

\appendix 

\section{Radial integrals}
\label{app:radial}

The radial integrals appearing in Eq.~(\ref{zeta}) are defined as 
\ba
U^\alpha_{j} &=& \langle \psi_{\rm gs} | \hat{U}_j | \psi_{\alpha} \rangle ~,
\nonumber\\
T^\alpha_{j,\mu} &=& \langle \psi_{\rm gs} | \hat{T}^\alpha_{j,\mu} | \psi_{\alpha} \rangle ~,
\ea
with $\mu=z$, $\pm$. Here $\alpha={\rm gs}$ for ground state baryons and $\alpha=\rho$ ($\lambda$) for $P$-wave excited baryons of the $\rho$-mode ($\lambda$-mode). For harmonic oscillator wave functions the radial integrals can be obtained in closed analytic form. The matrix elements, $U^\alpha_{j}$ and $T^\alpha_{j,-}$, which are relevant for the analysis of electromagnetic couplings were already given in Ref.~\cite{EOP2023}. For the sake of completeness we present here the results for all matrix elements. 

\noindent
(i) gs $\rightarrow$ gs:
\ba
U^{\rm gs}_1(k) &=& \mbox{e}^{-k^2/8\alpha_{\rho}^2} \, 
\mbox{e}^{-(\frac{3m'}{2m+m'} k)^2/24\alpha_{\lambda}^2} ~,
\nonumber\\
U^{\rm gs}_2(k) &=& U^{\rm gs}_1(k) ~, 
\nonumber\\
U^{\rm gs}_3(k) &=& \mbox{e}^{-(\frac{3m}{2m+m'} k)^2/6\alpha_{\lambda}^2} ~,
\ea
and
\begin{align}
T^{\rm gs}_{1,z}(k)
 &= -mk_0k
 \left[
   \frac{1}{4\alpha_{\rho}^{2}}
   +\left(\frac{3m'}{2m+m'}\right)^{2}
    \frac{1}{12\alpha_{\lambda}^{2}}
 \right]
 \notag\\[-1mm]
 &\quad\times
 \exp\!\left(-\frac{k^{2}}{8\alpha_{\rho}^{2}}\right)
 \notag\\[-1mm]
 &\quad\times
 \exp\!\left[-\frac{1}{24\alpha_{\lambda}^{2}}
 \left(\frac{3m'k}{2m+m'}\right)^{2}\right],
 \notag\\
T^{\rm gs}_{2,z}(k)
 &= T^{\rm gs}_{1,z}(k),
 \notag\\
T^{\rm gs}_{3,z}(k)
 &= -\frac{m'k_0k}{3\alpha_{\lambda}^{2}}
 \left(\frac{3m}{2m+m'}\right)^{2}
 \notag\\[-1mm]
 &\quad\times
 \exp\!\left[-\frac{1}{6\alpha_{\lambda}^{2}}
 \left(\frac{3mk}{2m+m'}\right)^{2}\right].
\end{align}

\noindent 
(i) $\rho \rightarrow$ gs:
\ba
U^{\rho}_1(k) &=& -i \frac{k}{2\alpha_{\rho}} \, 
\mbox{e}^{-k^2/8\alpha_{\rho}^2} \, 
\mbox{e}^{-(\frac{3m'}{2m+m'} k)^2/24\alpha_{\lambda}^2} ~,
\nonumber\\
U^{\rho}_2(k) &=& -U^{\rho}_1(k) ~, 
\nonumber\\ 
U^{\rho}_3(k) &=& 0 ~,
\ea
and
\begin{align}
T^{\rho}_{1,z}(k)
 &= -i\frac{mk_0}{2\alpha_{\rho}}
 \exp\!\left(-\frac{k^{2}}{8\alpha_{\rho}^{2}}\right)
 \notag\\[-1mm]
 &\quad\times
 \exp\!\left[-\frac{1}{24\alpha_{\lambda}^{2}}
 \left(\frac{3m'k}{2m+m'}\right)^{2}\right]
 \notag\\[-1mm]
 &\quad\times\Biggl[
 1-\frac{k^{2}}{4\alpha_{\rho}^{2}}
 \notag\\[-1mm]
 &\hspace{17mm}{}
 -\left(\frac{3m'}{2m+m'}\right)^{2}
  \frac{k^{2}}{12\alpha_{\lambda}^{2}}
 \Biggr],
 \notag\\
T^{\rho}_{1,\pm}(k)
 &= \mp i\frac{mk_0}{\sqrt{2}\alpha_{\rho}}
 \exp\!\left(-\frac{k^{2}}{8\alpha_{\rho}^{2}}\right)
 \notag\\[-1mm]
 &\quad\times
 \exp\!\left[-\frac{1}{24\alpha_{\lambda}^{2}}
 \left(\frac{3m'k}{2m+m'}\right)^{2}\right],
 \notag\\
T^{\rho}_{2,\mu}(k)
 &= -T^{\rho}_{1,\mu}(k),
 \notag\\
T^{\rho}_{3,\mu}(k)
 &= 0.
\end{align}

\noindent 
(i) $\lambda \rightarrow$ gs:
\begin{align}
U^{\lambda}_{1}(k)
 &= -i\frac{k}{2\sqrt{3}\alpha_{\lambda}}
 \frac{3m'}{2m+m'}
 \notag\\[-1mm]
 &\quad\times
 \exp\!\left(-\frac{k^{2}}{8\alpha_{\rho}^{2}}\right)
 \notag\\[-1mm]
 &\quad\times
 \exp\!\left[-\frac{1}{24\alpha_{\lambda}^{2}}
 \left(\frac{3m'k}{2m+m'}\right)^{2}\right],
 \notag\\
U^{\lambda}_{2}(k)
 &= U^{\lambda}_{1}(k),
 \notag\\
U^{\lambda}_{3}(k)
 &= i\frac{k}{\sqrt{3}\alpha_{\lambda}}
 \frac{3m}{2m+m'}
 \notag\\[-1mm]
 &\quad\times
 \exp\!\left[-\frac{1}{6\alpha_{\lambda}^{2}}
 \left(\frac{3mk}{2m+m'}\right)^{2}\right].
\label{ulambda}
\end{align}
and
\begin{align}
T^{\lambda}_{1,z}(k)
 &= -i\frac{mk_0}{2\sqrt{3}\alpha_{\lambda}}
 \frac{3m'}{2m+m'}
 \notag\\[-1mm]
 &\quad\times
 \exp\!\left(-\frac{k^{2}}{8\alpha_{\rho}^{2}}\right)
 \notag\\[-1mm]
 &\quad\times
 \exp\!\left[-\frac{1}{24\alpha_{\lambda}^{2}}
 \left(\frac{3m'k}{2m+m'}\right)^{2}\right]
 \notag\\[-1mm]
 &\quad\times\Biggl[
 1-\frac{k^{2}}{4\alpha_{\rho}^{2}}
 \notag\\[-1mm]
 &\hspace{17mm}{}
 -\left(\frac{3m'}{2m+m'}\right)^{2}
  \frac{k^{2}}{12\alpha_{\lambda}^{2}}
 \Biggr],
 \notag\\
T^{\lambda}_{3,z}(k)
 &= i\frac{m'k_0}{\sqrt{3}\alpha_{\lambda}}
 \frac{3m}{2m+m'}
 \notag\\[-1mm]
 &\quad\times
 \exp\!\left[-\frac{1}{6\alpha_{\lambda}^{2}}
 \left(\frac{3mk}{2m+m'}\right)^{2}\right]
 \notag\\[-1mm]
 &\quad\times\Biggl[
 1-\left(\frac{3m}{2m+m'}\right)^{2}
 \notag\\[-1mm]
 &\hspace{27mm}{}
 \times\frac{k^{2}}{3\alpha_{\lambda}^{2}}
 \Biggr],
 \notag\\
T^{\lambda}_{1,\pm}(k)
 &= \mp i\frac{mk_0}{\sqrt{6}\alpha_{\lambda}}
 \frac{3m'}{2m+m'}
 \notag\\[-1mm]
 &\quad\times
 \exp\!\left(-\frac{k^{2}}{8\alpha_{\rho}^{2}}\right)
 \notag\\[-1mm]
 &\quad\times
 \exp\!\left[-\frac{1}{24\alpha_{\lambda}^{2}}
 \left(\frac{3m'k}{2m+m'}\right)^{2}\right],
 \notag\\
T^{\lambda}_{3,\pm}(k)
 &= \pm i\sqrt{\frac{2}{3}}\,
 \frac{m'k_0}{\alpha_{\lambda}}
 \frac{3m}{2m+m'}
 \notag\\[-1mm]
 &\quad\times
 \exp\!\left[-\frac{1}{6\alpha_{\lambda}^{2}}
 \left(\frac{3mk}{2m+m'}\right)^{2}\right],
 \notag\\
T^{\lambda}_{2,\mu}(k)
 &= T^{\lambda}_{1,\mu}(k).
\label{tlambda}
\end{align}
The values of the harmonic oscillator size parameters, $\alpha_\rho$ and $\alpha_\lambda$, are taken from Table~VIII of Ref.~\cite{EOP2023}. For the sake of completeness, the values are presented in Table~\ref{alphas}.

\begin{table}[t]
\centering
\caption{Values of oscillator size parameters in MeV.}
\label{alphas}
\begin{tabular}{ccccc}
\noalign{\smallskip}
\toprule
& \multicolumn{2}{c}{$Q=c$} & \multicolumn{2}{c}{$Q=b$} \\
\noalign{\smallskip}
\cmidrule(lr){2-3} \cmidrule(lr){4-5}	 
\noalign{\smallskip}
& $\alpha_\rho$ & $\alpha_\lambda$ & $\alpha_\rho$ & $\alpha_\lambda$ \\
\noalign{\smallskip}
\hline
\noalign{\smallskip}
$\Sigma_Q$, $\Lambda_Q$ & 392 & 478 & 380 & 486 \\
$\Xi_Q$, $\Xi'_Q$       & 418 & 500 & 405 & 514 \\
$\Omega_Q$              & 440 & 517 & 426 & 537 \\
$\Xi_{QQ}$              & 601 & 425 & 771 & 417 \\
$\Omega_{QQ}$           & 601 & 471 & 771 & 466 \\
\noalign{\smallskip}
\botrule
\end{tabular}
\end{table}

\section{Baryon flavor wave functions}
\label{app:wf}

In this section we present the baryon flavor wave functions using the Baird-Biedenharn phase convention for the $SU(3)$ flavor couplings \cite{Baird:1963wv}. We only show the highest charge state $I_3=I$ with $Q=I+Y/2$. The other charge states are obtained by applying the lowering operator in isospin space.

\noindent
(i) Octet $qqq$ baryons, $(p,q)=(1,1)$ 
\ba
p_{\rho} &:& (udu - duu)/\sqrt{2} ~, 
\nonumber\\
\Sigma^+_{\rho} &:& (usu - suu)/\sqrt{2} ~, 
\nonumber\\
\Lambda_{\rho} &:& \frac{1}{\sqrt{12}}\bigl(2uds-2dus-dsu
\nonumber\\
&& \qquad {}+sdu-sud+usd\bigr) ~,
\nonumber\\
\Xi^0_{\rho} &:& (uss - sus)/\sqrt{2} ~.
\ea
and
\ba
p_{\lambda} &:& (2uud - udu - duu)/\sqrt{6} ~, 
\nonumber\\
\Sigma^+_{\lambda} &:& (2uus - usu - suu)/\sqrt{6} ~, 
\nonumber\\
\Lambda_{\lambda} &:& (- dsu - sdu + sud + usd)/2 ~, 
\nonumber\\
\Xi^0_{\lambda} &:& (sus + uss - 2ssu)/\sqrt{6} ~.
\ea

\noindent
(ii) Decuplet $qqq$ baryons, $(p,q)=(3,0)$
\ba
\Delta^{++} &:& uuu ~, 
\nonumber\\
\Sigma^{+} &:& (suu + usu + uus)/\sqrt{3} ~, 
\nonumber\\
\Xi^{0} &:& (ssu + sus + uss)/\sqrt{3} ~, 
\nonumber\\
\Omega^{-} &:& sss ~. 
\ea

\noindent
(iii) Singlet $qqq$ baryon, $(p,q)=(0,0)$
\begin{equation}
\begin{aligned}
\Lambda^\ast :{}&\quad
\frac{1}{\sqrt{6}}\bigl(
 uds-dus+dsu-sdu \\
&\hspace{24mm}{}+sud-usd
\bigr).
\end{aligned}
\end{equation}

\noindent
(iv) Sextet $qqQ$ baryons, $(p,q)=(2,0)$
\ba
\Sigma_Q &:& uuQ ~,
\nonumber\\
\Xi'_Q &:& (us+su)Q/\sqrt{2} ~,
\nonumber\\
\Omega_Q &:& ssQ ~.
\ea

\noindent
(v) Anti-triplet $qqQ$ baryons, $(p,q)=(0,1)$
\ba
\Lambda_Q &:& (ud-du)Q/\sqrt{2} ~,
\nonumber\\
\Xi_Q &:& (us-su)Q/\sqrt{2} ~.
\ea 

\noindent
(vi) Triplet $QQq$ baryons, $(p,q)=(1,0)$
\ba
\Xi_{QQ} &:& QQu ~,
\nonumber\\
\Omega_{QQ} &:& QQs ~.
\ea

\section{Meson flavor operators}
\label{app:meson}

In this appendix we present the flavor operator $X^M_j$ for the emission of a meson $M$ from the $j$-th quark using the Baird-Biedenharn phase convention for the $SU(3)$ flavor couplings \cite{Baird:1963wv}. 
For light mesons the flavor operators are proportional to the Gell-Mann matrices \cite{PhysRevD.21.1868,PhysRevD.55.2862,BIJKER200089}. 

\noindent
(i) Singlet meson, $(p,q)=(0,0)$
\ba
X^{\eta_1} \;=\; \sqrt{\frac{2}{3}} {\cal I} \;=\; (u\bar{u}+d\bar{d}+s\bar{s}) \sqrt{2/3} ~.
\ea

\noindent
(ii) Octet mesons, $(p,q)=(1,1)$
\ba
X^{K^0} \;=& (\lambda_6 - i\lambda_7)/\sqrt{2} &=\; d\bar{s} \sqrt{2} ~,
\nonumber\\
X^{K^+} \;=& (\lambda_4 - i\lambda_5)/\sqrt{2} &=\; u\bar{s} \sqrt{2} ~,
\nonumber\\
X^{\pi^-} \;=& (\lambda_1 + i\lambda_2)/\sqrt{2} &=\; d\bar{u} \sqrt{2} ~,
\nonumber\\
X^{\pi^0} \;=& \lambda_3 &=\; u\bar{u}-d\bar{d} ~,
\nonumber\\
X^{\pi^+} \;=& -(\lambda_1 - i\lambda_2)/\sqrt{2} &=\; -u\bar{d} \sqrt{2} ~,
\nonumber\\
X^{\eta_8} \;=& -\lambda_8 \\
&=& -\frac{u\bar{u}+d\bar{d}-2s\bar{s}}{\sqrt{3}} ~,
\nonumber\\
X^{K^-} \;=& (\lambda_4 + i\lambda_5)/\sqrt{2} &=\; s\bar{u} \sqrt{2} ~,
\nonumber\\
X^{\overline{K}^0} \;=& -(\lambda_6 + i\lambda_7)/\sqrt{2} &=\; -s\bar{d} \sqrt{2} ~.
\ea


\backmatter
\section*{Statements and Declarations}

\bmhead{Competing interests}
The authors declare that they have no competing interests.

\bmhead{Data availability}
No datasets were generated or analysed during the current study.

\ifuseprebuiltbbl
  \IfFileExists{main.bbl}{%
  }{%
    \PackageError{EPJC patch}{main.bbl was not found}%
      {Upload the main.bbl from the compiling PRD project, or set
       \string\useprebuiltbblfalse and repair biblio.bib.}%
  }
\else
    \nocite{*}
    \bibliography{maino}
\fi


\end{document}